\documentclass[journal]{IEEEtran}

\usepackage{lineno,hyperref}

\usepackage{hyperref}
\hypersetup{colorlinks,allcolors=black}
\usepackage{multicol}
\usepackage{bm}
\usepackage{mathrsfs}
\usepackage{amsmath}
\usepackage{amssymb}
\usepackage{graphicx}
\usepackage{subfigure}
\usepackage{stfloats}
\usepackage{float}
\usepackage{multirow}
\usepackage{epstopdf}
\usepackage{cite}
\usepackage{color}
\modulolinenumbers[5]
\ifCLASSINFOpdf
\else
\fi
\begin{document}
	
\title{Real-time LCC-HVDC Maximum Emergency Power Capacity Estimation Based on Local PMU Measurements}

\author{Long Peng,~\IEEEmembership{Student Member,~IEEE}, Junbo Zhao,~\IEEEmembership{Senior Member,~IEEE}, Yong Tang,~\IEEEmembership{Senior Member,~IEEE}, Lamine Mili, \IEEEmembership{Life Fellow,~IEEE}, Zhuoyuan Gu,~\IEEEmembership{Member,~IEEE}, Zongsheng Zheng,~\IEEEmembership{Student Member,~IEEE}
	
\thanks{This work was supported by National Key Research and Development Program of China (2016YFB0900600) and Technology Projects of State Grid Corporation of China (52094017000W). L. Mili was supported by the US National Science Foundation under grant ECCS 1711191.}

\thanks{L. Peng, Y. Tang and Z. Gu are with China Electrical Power Research Institute. (e-mail: penglong\_epri@163.com; tangyong@epri.sgcc.com).}

\thanks{J. Zhao is with the Department of Electrical and Computer Engineering, Mississippi State University, Starkville, MS 39762 USA (e-mail: junbo@ece.msstate.edu).}

\thanks{L. Mili is with the Bradley Department of Electrical and Computer Engineering, Virginia Polytechnic Institute and State University, Falls Church, VA 22043 USA (e-mail: lmili@vt.edu).}

\thanks{Z. Zheng is with the School of Electrical Engineering, Southwest Jiaotong University, Chengdu 611756, China (e-mail: bk20095185@my.swjtu.edu.cn).}}

\markboth{IEEE TRANSACTIONS ON POWER SYSTEMS, 2019}%
{Shell \MakeLowercase{\textit{et al.}}: Bare Demo of IEEEtran.cls for IEEE Journals}

\maketitle
\begin{abstract}
The adjustable capacity of a line-commutated-converter High Voltage Direct Current (LCC-HVDC) connected to a power system, called the LCC-HVDC maximum emergency power capability or HVDC-MC for short, plays an important role in determining the response of that system to a large disturbance. However, it is a challenging task to obtain an accurate HVDC-MC due to system model uncertainties as well as to contingencies. To address this problem, this paper proposes to estimate the HVDC-MC using a Thevenin equivalent (TE) of the system seen at the HVDC terminal bus of connection with the power system, whose parameters are estimated by processing positive-sequences voltages and currents of local synchrophasor measurements.
{{The impacts of TE potential changes on the impedance estimation under large disturbance have been extensively investigated and an adaptive screening process of current measurements is developed to reduce the error of TE impedance estimation}}. The uncertainties of phasor measurements have been further taken into account by resorting to the total least square estimation method. {{The limitations of the HVDC control characteristics, the voltage dependent current order limit, the converter capacity and the AC voltage on HVDC-MC estimation are also considered.}} The simulations show that the proposed method can accurately track the dynamics of the TE parameters and the real-time HVDC-MC after the large disturbances.
\end{abstract}
\begin{IEEEkeywords}
Parameter estimation, HVDC, Emergency control, Thevenin equivalent, PMU.
\end{IEEEkeywords}
	
\IEEEpeerreviewmaketitle
	
\section{Introduction}

\IEEEPARstart
{H}{VDC} transmission technology has been widely used in today's power systems due to its capability of transmitting large-capacity power over a long-distance, with less losses {{as compared to the AC transmission technology \cite{benasla2018hvdc}. According to \cite{alassi2019hvdc}, LCC-HVDC projects still dominate the HVDC market of high power and voltage ratings, which can be up to $8\sim12$ $GW$ with voltages up to $800\sim1100$ $kV$. 
With the increased penetration of distributed energy resources and the formation of multi-HVDC systems, 
%small malfunctions can cause tripping of large wind/solar farms or a continuous commutation failure of multi-HVDCs 
it is easy to cause tripping of large wind/solar farms or a continuous commutation failure of multi-HVDC \cite{sun2017renewable,shao2017fast}.
%With the reduced system inertia due to the increased penetration of renewable energy resources, it is more and more challenging to deal with system disturbances, such as tripping of large wind/solar farms or a continuous commutation failure of multi-HVDCs \cite{sun2017renewable,shao2017fast}. 
These faults can lead to large power imbalance in a short period, and subsequently the frequency or angle instability \cite{elizondo2018interarea,elizondo2017hvdc}. 
%Under the normal operation, it is generally desirable to keep the HVDC at a fixed power to avoid negative interactions. However, HVDC can also support the AC system to quickly recover to normal condition without inducing stability issue to HVDC.
Compared with other emergency measures, such as generator tripping and load shedding, HVDC emergency power modulation has the advantages of fast control without loss of components, so it is a good candidate for emergency control\cite{benasla2019power}.
%However, as the commutation relies on AC system voltages, its failures can occur in the presence of large voltage disturbances, especially for the multi-HVDC systems when a large power imbalance exists. Thus, a fast power support from different HVDCs can be a good candidate for emergency control. 
To achieve the control goal, an accurate estimation of the maximum emergency power capability of the HVDC must be determined first. Otherwise, the control may not be sufficient for maintaining system stability \cite{liu2017design}. To address this problem, this paper develops a synchrophasor measurements-based method for the HVDC-MC estimation.}} 

In the literature, the main HVDC emergency control strategies are derived from off-line {{planning}}/simulations, where the strategy table is formed based on the model and the expected fault set. In this way, when the emergent fault is detected in the off-line strategy table, the corresponding scheme can be adopted for the emergency control \cite{qi2010research,gomez2011prediction}. However, with the integration of more and more inverter-based distributed energy resources and flexible loads, it becomes a challenge to obtain accurate system models and parameters. Therefore, the simulated case may not match reality \cite{xu2018propagating}. In addition, there are many combinations of faults and the investigations of all of them are not very realistic. {{As a result, the off-line planning strategy may miss several scenarios, especially the low-probability but high-risk cascading faults, which may lead to instability or even blackouts \cite{andersson2005causes}.
Furthermore, the planning strategy is typically derived for N-1 security criterion and for the worst estimated scenario, which may be too conservative in some scenarios \cite{makarov2012pmu}.}}
%Note that the blackouts involve cascading reactions with multiple faults, which are very time-consuming to simulate under different scenarios 
% Hence, the traditional methods {{based on off-line simulations}} have faced great difficulties in managing the risk 
{{In summary, the emergency control strategy based on the off-line planning/simulations is not adaptive to the varying operation conditions, which could result in excessive or deficient actions.}}
%especialy under the low probability high impact cascading events.
Thanks to the widespread development of phasor measurement units (PMU)\cite{kamwa2006wide,zhao2019power}, the real-time monitoring of HVDC-MC becomes possible. 
%{{In case of a large disturbance, the power and generation unbalance can be quite large, which requires fast control actions that HVDCs can provide. 
Note that the HVDC-MC is affected by the voltage support capability of the AC system.
%If the latter is weak, the decreased AC voltage will limit the HVDC power outputs  \cite{krishayya1997ieee,balu1994power}}}. 
%This is because when DC current reaches a certain point, the further increase of the current will reduce DC power output. In other words, the increased consumption in reactive power causes both AC and DC voltages to drop and the decline in DC voltage is greater than the increase in DC current. 
If the latter is weak, the increase of the DC current will cause both AC and DC voltages to a significant drop, which will limit the HVDC power outputs \cite{krishayya1997ieee,balu1994power}.
%When DC current reaches a certain point, the further increase of the current will reduce DC power output.This is because the increased consumption in reactive power causes both AC and DC voltages to drop and the decline in DC voltage is greater than the increase in DC current.
To quantify the system's voltage support capability, a measurement-based Thevenin equivalent (TE) can be developed. %thanks to the widespread development of phasor measurement units (PMU) \cite{makarov2012pmu,kamwa2006wide,zhao2019power}.}}
By using the PMU measurements, the TE parameters of the AC system can be tracked and therefore the HVDC-MC can be estimated considering the AC/DC model and operational constraints.

%{{The TE parameters can be obtained based on the the power grid topology. However, the accuracy of these information is difficult to guarantee for real-time application \cite{ajjarapu1992continuation,yun2019online}.
{{The TE parameters can be obtained based on the the concrete power grid topology \cite{yun2019online}. However, the reliance on the whole network information increases the difficulty of real-time application. To this end, the methods based on local measurements have been proposed.}} 
%{{The TE parameters depends on power grid topology and its parameters \cite{ajjarapu1992continuation,yun2019online}. However, the accuracy of these information is difficult to guarantee for real-time application. To this end, local-measurement-based methods have been proposed.}} 
In \cite{vu1999use}, the least squares method was used to estimate the potential and impedance based on voltage and current measurements. This approach was further enhanced to track the change of the TE parameters utilizing a time window or the recursive least squares with a forgetting factor \cite{wang2012real,chen2014pmu,babazadeh2017real}. {{In \cite{zhao2016robust}, a robust recursive least squares estimation method was proposed to handle the gross errors in the voltage measurements. In \cite{su2018robust}, the second-order cone programming technique was used to address uncertainties in the current measurements. These methods assume that the magnitude and phase angle of TE potential in the observation window are unchanged. In \cite{abdelkader2014online}, the influence of system side change on the estimation of TE parameters was investigated and a method was proposed that can handle phase angle shift. However, it cannot deal with the variation of TE potential magnitude. In addition, all the above methods are used for the static voltage analysis assuming continuous load increase. 
To the best of our knowledge, TE estimation with large disturbances is still an open problem.
%After the large disturbance, the TE impedance will change according to the system topology, but it can remain unchanged at least for a short period {{due to the discrete characteristic of the topology change. However, 
%%it should be noted that 
%the characteristics of the load-side and TE potential are quite different from the continuous load increase. 
Specifically, 
%the load current does not keep changing for a long period as compared to the continuous load increase. 
%Different from the case of continuously increasing load, 
after a large disturbance, the load will initially have a large voltage and current change, but will gradually return to a stable state.
When the system is stabilized again, the TE parameters are not observable because the measurements are invariant. That means not all measurements can be used.
%TE parameters are not observable based on the local measurements under stable state.
%the measurements with little or no change in the current must be screened out, otherwise the observability cannot be guaranteed.
Furthermore,
both the magnitude and the phase angle of the TE potential are changing, and the change is typically larger than that of static voltage analysis scenarios without large disturbances. So the effect of potential change on the TE estimation should be fully taken into account.
%Therefore, the existing methods could not handle the scenario of large disturbances.
%The existing methods are typically used for static analysis. The key insight is that the current measurements must be carefully chosen to ensure the observability of the TE parameters.
}}
\begin{figure*}[b]
	\centering
	\includegraphics[scale=0.45] {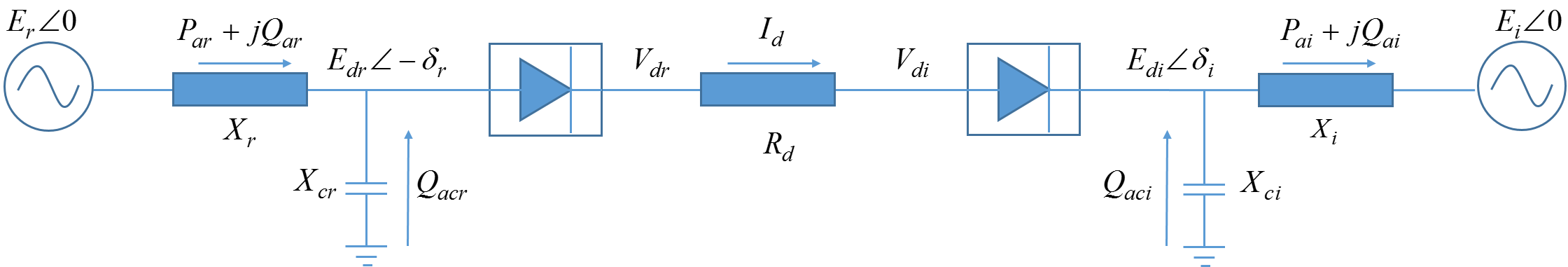}
	\caption{Equivalent structure of the AC/DC system. It consists of two generators connected via a HVDC link}
	\label{1}
\end{figure*}

To address the aforementioned issues, this paper proposes a synchrophasor measurements-driven estimation method that is able to track TE parameters in the presence of large disturbance and yield an accurate HVDC-MC. A sensitivity analysis is first carried out to assess the error of the TE impedance estimation caused by the system potential variations. Based on the system dynamic regulation characteristics, the upper bound of the TE potential variation in the presence of large disturbance can be determined. 
This allows us to develop adaptive measurement screening strategy for enhancing the observability of TE parameters.
To filter out the synchrophasor measurement noise, the total least squares (TLS) method is used. To this end, the HVDC-MC estimation considering various regulation and operational constraints can be obtained to inform emergency control. The developed robust TE estimation method is general and not restricted to LCC-HVDC system.
%{{To overcome the aforementioned drawbacks, this paper proposes a synchrophasor measurements-driven estimation method that is able to track TE parameters during the large disturbance and yield an accurate HVDC-MC. A sensitivity analysis is first carried out to assess the error of the TE impedance estimation caused by the system potential variations. This allows us to determine the upper bound of the TE potential variation in the presence of large disturbance, and to develop adaptive measurement screening strategy for enhancing the observability of TE parameters. To filter out the synchrophasor measurement noise, the total least squares (TLS) method is developed. To this end, the HVDC-MC estimation considering various regulation and operational constraints can be obtained to inform emergency control. To the best of our knowledge, TE estimation with large disturbances is still an open problem. The existing methods are typically used for static analysis. The developed robust TE estimation method is general and not restricted to LCC-HVDC system.}}

The rest of the paper is organized as follows. {{In Section \uppercase\expandafter{\romannumeral2}, the equivalent model of the AC/DC hybrid system and the constraints for HVDC-MC are introduced. In Section \uppercase\expandafter{\romannumeral3}, the TE estimation method considering the variation of TE potential is proposed. In Section \uppercase\expandafter{\romannumeral4}, the threshold for selecting current measurements is derived and the detailed algorithm is presented. Numerical results are conducted and analyzed in Section \uppercase\expandafter{\romannumeral5}. Finally, Section \uppercase\expandafter{\romannumeral6} concludes the paper.}} 

\section{AC/DC Hybrid System Modeling}
Fig. \ref{1} displays the equivalent structure of an AC/DC hybrid system power system. The main variables include the equivalent voltage $E\angle 0$ and the reactance $X$ (ignoring the resistance), the voltage ${{E}_{d}}\angle \delta $ of the AC buses at the converter station, the active power ${{P}_{a}}$ and the reactive power ${{Q}_{a}}$ of the AC line {{flowing into or out of the converter station}},  the reactance ${{X}_{c}}$ of the reactive power compensator that corresponds to the reactive power injection ${{Q}_{ac}}$, the DC voltage ${{V}_{d}}$, the DC current ${{I}_{d}}$, and the DC line resistance ${{R}_{d}}$. The subscript $r$ and $i$ represent the rectifier and inverter, respectively.

\vspace{-0.3cm}
\subsection{The AC System Model}

Power flow equations of the AC system are as follows:

\begin{equation}
{{P}_{ar}}=\frac{{{E}_{r}}{{E}_{dr}}}{{{X}_{r}}}\sin {{\delta }_{r}}, {{Q}_{ar}}=\frac{{{E}_{r}}{{E}_{dr}}\cos {{\delta }_{r}}-E_{dr}^{2}}{{{X}_{r}}},
\end{equation}

\begin{equation}
{{P}_{ai}}=\frac{{{E}_{i}}{{E}_{di}}}{{{X}_{i}}}\sin {{\delta }_{i}}, {{Q}_{ai}}=\frac{E_{di}^{2}-{{E}_{i}}{{E}_{di}}\cos {{\delta }_{i}}}{{{X}_{i}}},
\end{equation}

\begin{equation}
{{Q}_{acr}}=\frac{E_{dr}^{2}}{{{X}_{cr}}}, {{Q}_{aci}}=\frac{E_{di}^{2}}{{{X}_{ci}}}
\label{e3}.
\end{equation}

According to the above equations, an upper voltage solution can be obtained as follows:

\begin{equation}
E_{dr}^{2}=\frac{\left( E_{r}^{2}-2{{Q}_{ar}}{{X}_{r}} \right)+\sqrt{{M}_{r}}}{2}
\label{e4},
\end{equation}
\begin{equation}
E_{di}^{2}=\frac{\left( E_{i}^{2}+2{{Q}_{ai}}{{X}_{i}} \right)+\sqrt{{M}_{i}}}{2}
\label{e5},
\end{equation}
where ${\ {M}_{r}={{\left( E_{r}^{2}-2{{Q}_{ar}}{{X}_{r}} \right)}^{2}}-4X_{ar}^{2}\left( P_{ar}^{2}+Q_{ar}^{2} \right)}$, ${\ {M}_{i}={{\left( E_{i}^{2}+2{{Q}_{ai}}{{X}_{i}} \right)}^{2}}-4X_{ai}^{2}\left( P_{ai}^{2}+Q_{ai}^{2} \right)}$.

\vspace{-0.3cm}
\subsection{The DC System Model}

The DC system can be represented by the following equations \cite{kundur1994power}:
\begin{equation}
{{V}_{dor}}=1.35{{B}_{r}}{{N}_{r}}{{E}_{dr}}, {{V}_{doi}}=1.35{{B}_{i}}{{N}_{i}}{{E}_{di}},
\label{e6}
\end{equation}
\begin{equation}
{{V}_{dr}}=\left( {{R}_{d}}-\frac{3}{\pi }{{B}_{i}}{{X}_{di}} \right){{I}_{d}}+{{V}_{doi}}\cos \gamma,
\label{e7}
\end{equation}
\begin{equation}
{{V}_{di}}={{V}_{doi}}\cos \gamma -\frac{3}{\pi }{{B}_{i}}{{X}_{di}}{{I}_{d}},
\label{e8}
\end{equation}
\begin{equation}
\cos \alpha =\frac{\frac{3}{\pi }{{B}_{r}}{{X}_{dr}}{{I}_{d}}+{{V}_{dr}}}{{{V}_{dor}}},
\label{e9}
\end{equation}
\begin{equation}
{{P}_{dr}}={{V}_{dr}}{{I}_{d}}, \cos {{\varphi }_{r}}={{{V}_{dr}}}/{{{V}_{dor}}},{{Q}_{dr}}={{P}_{dr}}\tan {{\varphi }_{r}},
\label{e10}
\end{equation}
\begin{equation}
{{P}_{di}}={{V}_{di}}{{I}_{d}}, \cos {{\varphi }_{i}}={{{V}_{di}}}/{{{V}_{doi}}}, {{Q}_{di}}={{P}_{di}}\tan {{\varphi }_{i}},
\label{e11}
\end{equation}
where ${{V}_{do}}$ is the ideal no-load direct voltage, $B$ is the number of bridges in series, $N$ is the transformer ratio, ${{X}_{d}}$ is the commutation reactance, $\gamma $ is the extinction angle, and $\alpha $ is the ignition angle, ${{P}_{d}}$ is active power , ${{Q}_{d}}$ is reactive power and $\varphi $ is the power factor angle.

\vspace{-0.3cm}
\subsection{Constraints of HVDC-MC}
The constraints of the control mode, VDCOL, the converter capacity and the AC voltage are presented below.
\subsubsection{Control Mode Constraint}
\begin{figure}[htb]
\setlength{\abovecaptionskip}{-0.1cm} 
    \centering 
    \subfigure[CD characteristic.] 
{ \label{CD} 
\includegraphics[width=0.4\columnwidth]{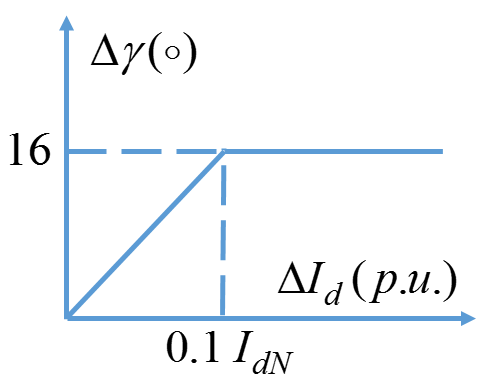}
} 
\subfigure[U-I curve of VDCOL] { \label{VDCOL} 
\includegraphics[width=0.45\columnwidth]{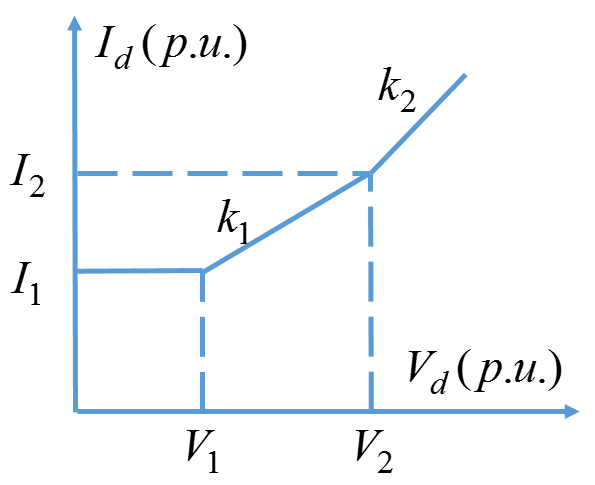}
} 
\caption{HVDC control characteristic curve}
	\label{23}
\end{figure}

in general, the rectifier has two main control modes, namely the constant current (CC) and the constant ignition angle (CIA). The inverter has the modes of constant current (CC), constant extinction angle (CEA) (or constant voltage) and current deviation (CD). According to the CIGRE HVDC benchmark system \cite{szechtman1991benchmark}, the following constraints in each mode are presented:

a) CC-CEA:
\begin{equation}
\alpha \ge {\alpha }_{min }, \gamma ={\gamma }_{min }, {{I}_{d}}={{I}_{ord}}, 
\label{CCCEA1}
\end{equation}

b) CIA-CD: 
\begin{equation}
\alpha ={{\alpha }_{min }}, \gamma ={{\gamma }_{min }}+\Delta \gamma , {{I}_{d}}={{I}_{ord}}-\Delta {{I}_{d}},
\end{equation}

c) CIA-CC: 
\begin{equation}
\alpha ={{\alpha }_{\min }}, {{I}_{d}}={{I}_{ord}}-{{I}_{m}},  
\label{CCCEA2}
\end{equation}
where ${\alpha}_{min}$ is the minimum ignition angle, ${\gamma}_{min} $ is the minimum extinction angle, ${I}_{ord}$ is the current order, the relationship between the current deviation $\Delta {{I}_{d}}$ and $\Delta \gamma$ is shown in Fig. \ref{CD} and $I_{dN}$ is the rated current, ${I}_{m}$ is the current margin between the rectifier and inverter, generally 0.1 $I_{dN}$.
Under normal operation, the rectifier controls the direct current with the CC mode while the inverter controls direct voltage with the CEA mode. As the power increase, the HVDC may operate in CIA-CD or CIA-CC. Since there exists the current deviation in the latter mode, the actual power cannot reach the power order. Note that in some practical projects, in order to compensate for the deviation caused by the control mode shift, the current margin compensation link is added. However, due to the large time constant of the link, it cannot rapidly increase the power outputs and thus it does not help much. So this paper considers CC-CEA as the main control mode during emergency power increase.
\vspace{-0.3cm}

\subsubsection{Voltage Dependent Current Order Limit (VDCOL) Constraint}
VDCOL can reduce the maximum allowable direct current when the voltage drops below a predetermined value. The current limitation ${I}_{VD}$ is shown in Fig. \ref{VDCOL} and (\ref{ee15}). VDCOL can be found in the rectifier and the inverter, and a certain current margin is maintained on each side.
\begin{equation}
{{I}_{VD}}=\left \{ \begin{aligned}
   & {{I}_{2}}+{{k}_{2}}({{V}_{d}}-{{V}_{2}}) & {{V}_{d}}\ge {{V}_{2}}  \\
   & {{I}_{1}}+{{k}_{1}}({{V}_{d}}-{{V}_{1}}) & {{V}_{1}}<{{V}_{d}}<{{V}_{2}}  \\
   & {{I}_{1}} & {{V}_{d}}\le {{V}_{1}}  \\
\end{aligned} .\right .
\label{ee15}
\end{equation}
\subsubsection{Converter Capacity Constraint}
the converter current constraint ${I}_{RA}$ is given by
\begin{equation}
{{I}_{RA}}=\left \{ \begin{aligned}
  & 1.3\sim1.5 {{I}_{dN}} & t\le 3 s \\ 
 & 1.05\sim1.1 {{I}_{dN}} & t>3 s \\ 
\end{aligned} .\right.
\label{ea15}
\end{equation}
where $t$ is the time. Adding the AC voltage constraints of $E_{min}$ and $E_{max}$, all the constraints during the power increase in CC-CEA mode are summarized as follows:
\begin{equation}
\begin{split}
 & \alpha \ge {{\alpha}_{min}}\\ 
 & I_d\le {{I}_{VD}}\\ 
 & I_d\le {{I}_{RA}}\\ 
 & E_{min}\le E_d \le E_{max}
\end{split}.
\label{constraint}
\end{equation}

Using the AC/DC model and the constraints above, we can calculate the HVDC-MC. It should be noted that the constraints can be modified according to the actual situation.
%especially for the angle constraints in control unit.

%{{\emph{Remark}: HVDC does not always run at the nominal power as it depends on the power demand. Moreover, for some weak system, HVDC may not run at the nominal power due to the N-1 or some other security stability constraint. In this paper, the HVDC maximum emergency power can be directly estimated based on local measurements when the model and constraints are determined.}}
%Monitoring the real-time available power is helpful for operators to master the operation and controllable state of HVDC.
%Furthermore, in the case of faults and if the AC-line is disconnected, HVDC is likely to be operated under non-nominal power operation. }}

\section{TE Estimation of AC System}
According to the AC and DC models shown in the last section, we know that the TE parameters of the AC system should be obtained first so as to calculate HVDC-MC. In this section, a synchrophasor-measurement-based method is proposed to achieve that goal.
%The sensitivity analysis to assess the error of the TE impedance estimation caused by the system potential variations is first performed. This provides us the way of determining the upper bound of the TE potential variation in the presence of large disturbance. Finally, the TLS method is formulated to filter PMU measurement noise.}}

\vspace{-0.3cm}
\subsection{Error Analysis of Impedance Estimation}
\begin{figure}[htb]
\setlength{\abovecaptionskip}{-0.1cm} 
	\centering
	\includegraphics[scale=0.6] {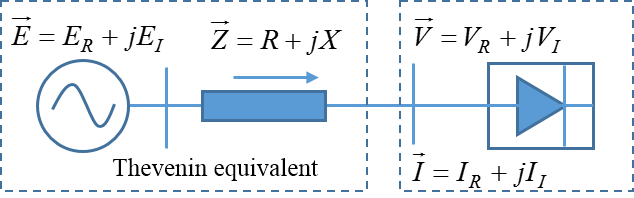}
	\caption{TE seen from the HVDC terminal bus}
	\label{2}
\end{figure}

%After a large disturbance, the HVDC will respond and recover quickly. The same thing is for the equivalent potential. In the following
The impact of the potential change on the impedance estimation is analyzed here. Fig. \ref{2} depicts a schematic diagram of the connection between the converter and the AC system. $\vec{E}$ is the TE potential, $\vec{Z}$ is the TE impedance, $\vec{V}$ is the bus voltage and $\vec{I}$ is the current. Then we have

\begin{equation}
\vec{E}-\vec{I}\cdot \vec{Z}=\vec{V}.
\label{e12}
\end{equation}
which can be further rewritten into the rectangular form as follows:
\begin{equation}
\left( \begin{matrix}
\begin{matrix}
1 & 0  \\
0 & 1  \\
\end{matrix} & \begin{matrix}
-{{I}_{R}} & {{I}_{I}}  \\
-{{I}_{I}} & -{{I}_{R}}  \\
\end{matrix}  \\
\end{matrix} \right)\left( \begin{matrix}
{{E}_{R}}  \\
{{E}_{I}}  \\
R  \\
X  \\
\end{matrix} \right)=\left( \begin{matrix}
{{V}_{R}}  \\
{{V}_{I}}  \\
\end{matrix} \right).
\label{ee19}
\end{equation}

If two PMU samples are used, the following equation can be derived:
\begin{equation}
-\vec{dI}\cdot \vec{Z}=\vec{dV}-\vec{dE},
\label{e14}
\end{equation}
where $\vec{dI}$, $\vec{dV}$ and $\vec{dE}$ represent the variations of current, voltage and TE potential at two samples, respectively. They can be further organized into the following form:
\begin{equation}
\left( \begin{matrix}
-d{{I}_{R}}  \\
-d{{I}_{I}}  \\
\end{matrix}\begin{matrix}
d{{I}_{I}}  \\
-d{{I}_{R}}  \\
\end{matrix} \right)\left( \begin{matrix}
R  \\
X  \\
\end{matrix} \right)=\left( \begin{matrix}
d{{V}_{R}}  \\
d{{V}_{I}}  \\
\end{matrix} \right)-\left( \begin{matrix}
d{{E}_{R}}  \\
d{{E}_{I}}  \\
\end{matrix} \right).
\label{e15}
\end{equation}
%where $dI_{Rn}$=$I_{Rn}$-$I_{Rn-1}$,$dI_{In}$=$I_{In}$-$I_{In-1}$,  %$dV_{Rn}$=$V_{Rn}$-$V_{Rn-1}$, $dV_{In}$=$V_{In}$-$V_{In-1}$.

From (\ref{e15}), $X$ and $R$ can be derived as follows:
\begin{equation}
X=\frac{d{{V}_{R}}d{{I}_{I}}-d{{V}_{I}}d{{I}_{R}}}{d{{I}_{R}}d{{I}_{R}}+d{{I}_{I}}d{{I}_{I}}}-\frac{d{{E}_{R}}d{{I}_{I}}-d{{E}_{I}}d{{I}_{R}}}{d{{I}_{R}}d{{I}_{R}}+d{{I}_{I}}d{{I}_{I}}},
\end{equation}
\begin{equation}
R=-\frac{d{{V}_{R}}d{{I}_{R}}+d{{I}_{I}}d{{V}_{I}}}{d{{I}_{R}}d{{I}_{R}}+d{{I}_{I}}d{{I}_{I}}}+\frac{d{{E}_{R}}d{{I}_{R}}+d{{I}_{I}}d{{E}_{I}}}{d{{I}_{R}}d{{I}_{R}}+d{{I}_{I}}d{{I}_{I}}}.
\end{equation}

They can be further written as
\begin{equation}
X={\bigl| \vec{dV}\times \vec{dI} \bigr|}/{{{\bigl| \vec{dI} \bigr|}^{2}}}-{\bigl| \vec{dE}\times \vec{dI} \bigr|}/{{{\bigl| \vec{dI} \bigr|}^{2}}},
\label{e18}
\end{equation}
\begin{equation}
R=-{\bigl| \vec{dV}\cdot \vec{dI} \bigr|}/{{{\bigl| \vec{dI} \bigr|}^{2}}}+{\bigl| \vec{dE}\cdot \vec{dI} \bigr|}/{{{\bigl| \vec{dI} \bigr|}^{2}}}.
\label{e19}
\end{equation}

The first term of  (\ref{e18}) and  (\ref{e19}) can be calculated with PMU measurements. Because $\vec{dE}$ is unknown, the second term cannot be obtained. If the first term is used as the estimated value, the estimation error of $X$ and $R$ are given by
\begin{equation}
e{{r}_{X}}={\bigl| \vec{dE}\times \vec{dI} \bigr|}/{{{\bigl| \vec{dI} \bigr|}^{2}}}={\bigl| \vec{dE} \bigr|\bigl| \sin {{\theta }_{EI}} \bigr|}/{\bigl| \vec{dI} \bigr|},
\end{equation}
\begin{equation}
e{{r}_{R}}=-{\bigl| \vec{dE}\cdot \vec{dI} \bigr|}/{{{\bigl| \vec{dI} \bigr|}^{2}}}=-{\bigl| \vec{dE} \bigr|\bigl| \cos {{\theta }_{EI}} \bigr|}/{\bigl| \vec{dI} \bigr|},
\end{equation}
where ${{\theta }_{EI}}$ is the angle between $\vec{dE}$ and $\vec{dI}$. Because $\bigl| \sin {{\theta }_{EI}} \bigr|<1$ and $\bigl| \cos {{\theta }_{EI}} \bigr|<1$, we know

\begin{equation}
\bigl| e{{r}_{X}} \bigr|<{\bigl| \vec{dE} \bigr|}/{\bigl| \vec{dI} \bigr|}, \bigl| e{{r}_{R}} \bigr|<{\bigl| \vec{dE} \bigr|}/{\bigl| \vec{dI} \bigr|}.
\label{er}
\end{equation}

To make sure the estimation error is acceptable, $\bigl| \vec{dE} \bigr|/\bigl| \vec{dI} \bigr|$ should be less than a certain value. To this end, $\bigl| \vec{dE} \bigr|$ should be small enough. This means that the sampling time interval should be as short as possible. Meanwhile, the measurements with little or no change in the currents must be screened out, otherwise the error would be very large. Furthermore, the current variation needs to be large enough so as to reduce the error caused by TE potential variation. The screening threshold of $\bigl| \vec{dI} \bigr|$ will be elaborated in Section \uppercase\expandafter{\romannumeral4}.

\vspace{-0.3cm}
\subsection{Parameter Estimation Using Total Least Squares}
The error of impedance estimation caused by potential change can be significantly reduced by fast sampling and selecting the measurements with larger $\bigl| \vec{dI} \bigr|$.  Considering model uncertainty and measurements noise, (\ref{e14}) can be written as
\begin{equation}
-\left( \vec{dI}-\vec{\varepsilon } \right)\cdot \vec{Z}+\vec{\eta }=\vec{dV},
\end{equation}
where $\vec{\varepsilon }$ represents the measurement error of current. The error of $\vec{Z}$ caused by $\vec{dE}$ and the measurement error of the voltage are expressed in $\vec{\eta }$. Due to the existence of model uncertainty and measurement noise, the least square method will lead to biased estimation results. To mitigate that, the TLS estimation is advocated in \cite{lavenius2015performance}.

Assume that $\vec{\varepsilon }$ and $\vec{\eta }$ obey the Gaussian distribution. By using the selected measurements in the time window of $ k $ intervals, we have the following equation at time $n$:
\begin{equation}
-\left( \begin{matrix}
{{{dI}}_{n-k+1}}-{{\vec{\varepsilon }}_{n-k+1}}  \\
{{{dI}}_{n-k+2}}-{{\vec{\varepsilon }}_{n-k+2}}  \\
\vdots   \\
{{{dI}}_{n}}-{{\vec{\varepsilon }}_{n}}  \\
\end{matrix} \right)\left( \begin{matrix}
R  \\
X  \\
\end{matrix} \right)
+\left( \begin{matrix}
{{\vec{\eta }}_{n-k+1}}  \\
{{\vec{\eta }}_{n-k+2}}  \\
\vdots   \\
{{\vec{\eta }}_{n}}  \\
\end{matrix} \right)=\left( \begin{matrix}
{{{dV}}_{n-k+1}}  \\
{{{dV}}_{n-k+2}}  \\
\vdots   \\
{{{dV}}_{n}}  \\
\end{matrix} \right)
\end{equation}
where ${{{dI}}_{n}}=\left( \begin{matrix}
d{{I}_{{{R}_{n}}}}  \\
d{{I}_{{{I}_{n}}}}  \\
\end{matrix}\begin{matrix}
-d{{I}_{{{I}_{n}}}}  \\
d{{I}_{{{R}_{n}}}}  \\
\end{matrix} \right)$, ${{{dV}}_{n}}=\left( \begin{matrix}
d{{V}_{{{R}_{n}}}}  \\
d{{V}_{{{I}_{n}}}}  \\
\end{matrix} \right)$.
Define the following two matrices:
\begin{equation}
A=-\left( \begin{matrix}
{{{dI}}_{n-k+1}}  \\
{{{dI}}_{n-k+2}}  \\
\vdots   \\
{{{dI}}_{n}}  \\
\end{matrix} \right), 
B=\left( \begin{matrix}
{{{dV}}_{n-k+1}}  \\
{{{dV}}_{n-k+2}}  \\
\vdots   \\
{{{dV}}_{n}}  \\
\end{matrix} \right).
\end{equation}

The TE impedance can be estimated by applying the singular value decomposition (SVD) as follows:

\begin{equation}
 \left( \begin{matrix}
A & B  \\
\end{matrix} \right)
=C\Lambda {{U}^{T}},
\end{equation}
where the columns of $C$ and $U$ are the left and right singular vectors, respectively. Then, the SVD computes only the first 3 columns of C via
\begin{equation}
 \left( \begin{matrix}
A & B  \\
\end{matrix} \right)
=\left( \begin{matrix}
{{C}_{1}} & {{C}_{2}}  \\
\end{matrix} \right) \begin{matrix}
\Sigma  \\
\end{matrix} \left( \begin{matrix}
U_{11} & U_{12}  \\
U_{21} & U_{22}  \\
\end{matrix} \right)^{T} \ ,
\end{equation}
where $C_1 \in \mathbb{R}^{2k \times 2}$, $C_2 \in \mathbb{R}^{2k \times 1}$, $\Sigma  \in \mathbb{R}^{3 \times 3}$, $ U_{11}  \in \mathbb{R}^{2 \times 2}$, $ U_{22}  \in \mathbb{R}^{1}$. Finally, the estimated values are shown in (\ref{err}), followed by the calculation of TE potential via (\ref{e12}). 
\begin{equation}
\hat{\left( \begin{matrix}
R  \\
X  \\
\end{matrix} \right)}=-{{U}_{12}}U_{22}^{-1}.
\label{err}
\end{equation}

\section{Algorithm of HVDC-MC Estimation}
%The measurements selection method for $\bigl| \vec{dI} \bigr|$ is derived and the detailed algorithm implementation of HVDC-MC is provided in this section.

%{{In the case of large disturbance, not all measurements can be used. The measurements with little or no change in the currents must be screened out, otherwise the observability cannot be guaranteed. Not only does the current variation need to be satisfied, but also the current variation needs to be large enough so as to reduce the error of impedance estimation caused by TE potential variation. Based on the analysis of $\bigl| \vec{dE} \bigr|$, 
This section develops an adaptive measurements selection method for $\bigl| \vec{dI} \bigr|$ to enhance TE parameters observability and the detailed algorithm implementation of HVDC-MC is also provided.

\vspace{-0.3cm}
\subsection{Proposed Measurements Selection Method}
{{Based on section \uppercase\expandafter{\romannumeral3}-A, the measurements with small $\bigl| \vec{dI} \bigr|$ should be screened out. To achieve that, a measurements selection method is proposed by analyzing the characteristics of $\bigl| \vec{dI} \bigr|$.}} The term $\bigl| \vec{dI} \bigr|$ after the large disturbance is analyzed to be divided into two main parts. One part is caused by the change of equivalent impedance of HVDC while the other part is caused by $\bigl| \vec{dE} \bigr|$. It should be noted that the excitation system, the angle oscillation and voltage regulators will affect the potential $\bigl| \vec{dE} \bigr|$. Because the time constant of the angle oscillation is larger than HVDC, $\bigl| \vec{dI} \bigr|$ will be dominated by $\bigl| \vec{dE} \bigr|$ after a certain time. At this time, $\bigl| \vec{dI} \bigr|$ is smaller, so the error term $\bigl| \vec{dE} \bigr|/\bigl| \vec{dI} \bigr|$ will become large or uncertain. As a result, all the measurements related to that should be screened out. 
Assume that when the HVDC returns into stable condition, the equivalent impedance $ \vec{{Z}_{d}}$ of the HVDC approximately remains unchanged. {{Then $\bigl| \vec{dI} \bigr|$ caused by $\bigl| \vec{dE} \bigr|$ is expressed as
 %\bigl| \vec{dE} \bigr|
\begin{equation}
 \bigl| \vec{dI} \bigr|=\frac{\bigl| \vec{dE} \bigr|}{\bigl| \vec{Z}+{{{\vec{Z}}}_{d}} \bigr|}= \frac{\bigl| \vec{dE} \bigr|}{\bigl| \vec{E} \bigr|}\bigl| \vec{I} \bigr| \leq \frac{{\bigl| \vec{dE} \bigr|}_{max}}{{\bigl| \vec{E} \bigr|}_{min}}\bigl| \vec{I} \bigr|,
 \end{equation}
 where ${\bigl| \vec{dE} \bigr|}_{max}$ denotes the upper bound of ${\bigl| \vec{dE} \bigr|}$ and ${\bigl| \vec{E} \bigr|}_{min}$ denotes the lower bound of ${\bigl| \vec{E} \bigr|}$. If we select measurements that satisfy
 \begin{equation}
 \bigl| \vec{dI} \bigr| > \frac{{\bigl| \vec{dE} \bigr|}_{max}}{{\bigl| \vec{E} \bigr|}_{min}}\bigl| \vec{I} \bigr|,
 \label{e36}
 \end{equation}
%the estimated values with the large errors can be first screened out, and the appropriate $\bigl| \vec{dI} \bigr|$ can be selected to further reduce the error.
the estimated values with the large or uncertainty errors can be first screened out. Then we can appropriately choose larger $\bigl| \vec{dI} \bigr|$ to further reduce the error.

From (\ref{e36}), ${{\bigl| \vec{dE} \bigr|}_{max}}$ and ${{\bigl| \vec{E} \bigr|}_{min}} $ are the key values to derive the threshold.
% The current $\bigl| \vec{I} \bigr|$ can be obtained from PMU. Then, in order to find the maximum $\bigl| \vec{dI} \bigr|$, 
% %caused only by the system change
% we need to analyze the lower bound of $\bigl| \vec{E} \bigr|$ and the upper bound of $\bigl| \vec{dE} \bigr|$. 
The latter can be obtained by substituting the possible values of TE reactance (neglecting the resistance) into (\ref{e12}), where $\vec{I}$ and $\vec{V}$ can be measured from PMUs. After the fault, the TE reactance typically does not decrease, so the minimum reactance can be set as the value when the network is fully connected. The maximum TE reactance can be determined by the worst topology case, for example in the presence of disconnecting several major lines. To simplify the derivation, this paper assumes ${{\bigl| \vec{E} \bigr|}_{min}} $ as 0.5 $p.u.$, which is typically a conservative lower bound of ${\bigl| \vec{E} \bigr|} $. 

For ${\bigl| \vec{dE} \bigr|}_{max}$, this paper mainly considers the influence of excitation system. We first analyze the maximum change of the potential of a single generator and then deduce the case of a multi-machine system.}} Specifically, under the no-load condition of the generator, the q-axis transient potential $ {{E}'}_{q} $ can be expressed as \cite{sauer1998power}:
\begin{equation}
{T_{ff}}\frac{{d{E_{qe}}}}{{dt}} + {E_{qe}} = {U_{ff}},  {T_{d0'}}\frac{{d{{E'}_q}}}{{dt}} + {{E'}_q} = {E_{qe}},
\label{29}
\end{equation}
where $T_{ff}$ is the exciter time constant, $E_{qe}$ is the no-load potential, $U_{ff}$ is input voltage of the exciter generator, $T_{d0'}$ is d-axis transient open-circuit time constant. Considering the maximum change of the generator internal potential, the input voltage variation of the exciter generator reaches the maximum value $ \Delta {U}_{\max } $ instantly after the large disturbance, yielding
 \begin{equation}
 \Delta {{{{E}'}}_{q}}(t)=\Delta {{U}_{\max }}\left( 1+\frac{{{e}^{-t/{{T}_{ff}}}}}{{{T}_{d{0}'}}/{{T}_{ff}}-1}+\frac{{{e}^{-t/{{T}_{d{0}'}}}}}{{{T}_{ff}}/{{T}_{d{0}'}}-1} \right),
 \label{35}
 \end{equation}
 where $\Delta {{{{E}'}}_{q}}(t) $ denotes the incremental form. The maximum variation of $ {{E}'}_q $ is given by
 \begin{equation}
 d{{{{E}'}_q}_{\max }}={{\left. \frac{d\Delta {{{{E}'}}_{q}(t)}}{dt} \right|}_{t={{t}_{0}}}}, {{t}_{0}}=\frac{{{T}_{d{0}'}}{{T}_{ff}}}{{{T}_{d{0}'}}-{{T}_{ff}}}\ln \frac{{{T}_{d{0}'}}}{{{T}_{ff}}}.
\label{32}
 \end{equation}
 For example, if $ {{T}_{ff}}  = 0.53\ s$, $ {{T}_{d{0}'}}  = 5\ s$, $ \Delta {{U}_{\max }}  = 10\ p.u. $, {{which is the maximum possible value according to the common excitation model parameters \cite{kundur1994power},}} so $ d{{{{E}'}_q}_{\max }}  =1.512 \ p.u./s $. During the sampling interval of 10 $ms$, the maximum change of $ {{{{E}'}}_{q}} $ is 0.01512 $ p.u. $ for a single generator.
 \vspace{-0.3cm} 
 \begin{figure}[htb]
 \setlength{\abovecaptionskip}{-0.1cm} 
 	\centering
 	\includegraphics[scale=0.7] {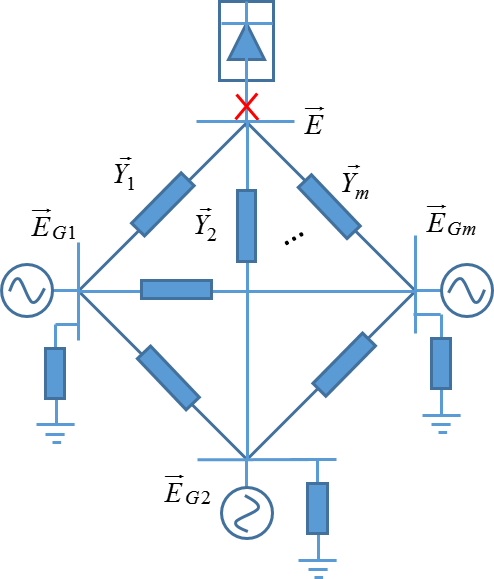}
 	\caption{Equivalent circuit of multi-machine with network reduction}
 	\label{fig3}
 \end{figure}
\vspace{-0.3cm}

{{We have found the maximum potential change corresponding to a single generator. Next, we analyze how the potential change of a single generator affects the TE potential.}}
For multi-machine system, the equivalent circuit that is reduced to generators and HVDC can be described by Fig. \ref{fig3}.
The TE potential $ \vec{E} $ is the voltage of the converter station when HVDC is operating as an open-circuit. Based on the Kirchhoff law, we have
\begin{equation}
\vec{E}(\vec{{Y}_{0}}+\sum\limits_{i=1}^{m}{ \vec{{Y}_{i}}})-\sum\limits_{i=1}^{m}{\vec{{{E}_{Gi}}}} \vec{{Y}_{i}} =0\ ,
\end{equation}
where $ \vec{{E}_{Gi}} $ is the generator potential, $\ m $ is the number of generators, $ \vec{{Y}_{i}} (G_i+jB_i) $ is the branch admittance, $ \vec{{Y}_{0}} (G_0+jB_0) $ is the  transfer admittance to ground at the HVDC terminal. Neglecting the transfer conductivity $G_i$ of the branch and the susceptance $B_0$ to ground, we get the following incremental form:
\begin{equation}
\vec{dE}({G}_{0}+\sum\limits_{i=1}^{m}{ {j{B}_{i}}})
=\sum\limits_{i=1}^{m}{ \vec{{{dE}_{Gi}}}} {{jB}_{i}}, 
\end{equation}
yielding
\begin{equation}
\bigl| \vec{dE} \bigr|=\frac{\bigl| \sum\limits_{i=1}^{m}{ \vec {dE}_{Gi} }  {jB}_{i} \bigr|} 
{\bigl|{G}_{0}+\sum\limits_{i=1}^{m}{  j{B}_{i} }\bigr|} 
\leq \frac{\sum\limits_{i=1}^{m}{ \bigl| \vec{ {dE}_{Gi} }  }  \bigr| \bigl| { {B}_{i} } \bigr|} 
{   \sum\limits_{i=1}^{m}{   \bigl|{B}_{i} \bigr|   }    }.
\end{equation}

\begin{table*}[t]
	\centering
	\caption{{{HVDC-MC Estimation ALGORITHM}}}
	\begin{tabular}{l l}
	\hline \hline
  \multicolumn{2}{l}{initialize: $k=20$, $\lambda$=0.8, A=null, B=null, ${I}_{d}$, $\Delta {I}_{d}$, $\gamma$=${\gamma}_{min}$, $E_{dr0}$, $E_{di0}$, $ \mu $=$10^{-4}$ } \\
	\hline
	\multicolumn{2}{l}{for $n=2,...$} \\
	\hline
	    1 & Obtain $V_{Rn}$, $V_{In}$, $I_{Rn}$ , $I_{In}$, $I_{dn}$. ($n$ starts from 1)	\\
		2 & $dV_{Rn}$=$V_{Rn}$-$V_{Rn-1}$, $dV_{In}$=$V_{In}$-$V_{In-1}$, $dI_{Rn}$=$I_{Rn}$-$I_{Rn-1}$, $dI_{In}$=$I_{In}$-$I_{In-1}$ \\ 
		%$\bigl| d{{{\vec{I}}}_{n}} \bigr|=\sqrt{dI_{Rn}^{2}+dI_{In}^{2}}$, $\bigl| {{{\vec{I}}}_{n}} \bigr|=\sqrt{I_{Rn}^{2}+I_{In}^{2}}$  \\
		3 & \textbf{if}  $\bigl| d{{{\vec{I}}}_{n}} \bigr|>0.1 \sim 0.2\bigl| {{\vec{I}}_{n}}  \bigr|$, $\textbf{then}$. \\
		4 & \qquad $ n_{0}=n$, $ \sigma =0.1 \sim 0.2\bigl| {{{\vec{I}}}_{n}} \bigr|\left( 1-{{\lambda }^{n-n_{0}}} \right).$\\
		5 & \textbf{end if}\\
	   6 &  $\textbf{if} \, \bigl| d{{\vec{I}}_{n}} \bigr|>\sigma$ ,  $0.5 p.u. < \bigl| {{\vec{E}}} \bigr|<1.5  p.u. $ , $ X>0 $ and  $ R >0 $, \textbf{then} \\
      7 & \qquad A = $\left( \begin{matrix}
A   \\
{ \begin{matrix}
-d{{I}_{Rn}}  \\
-d{{I}_{In}}  \\
\end{matrix}
\begin{matrix}
d{{I}_{In}}  \\
-d{{I}_{Rn}}  \\
\end{matrix} }
\end{matrix} \right)  
\text{and} \,
 B = {\left( \begin{matrix}
B   \\
{ \begin{matrix}
d{{V}_{Rn}}  \\
d{{V}_{In}}  \\
\end{matrix}}
\end{matrix}
 \right)}.\,\text{Keep the data only from $n-k+1$ to $n$. Find the SVD of (A B) and} \left( \begin{matrix}
R_{n}  \\
X_{n}  \\
\end{matrix} \right)=-{U}_{12}U_{22}^{-1}$. \\
		8 &  \textbf{else} \\
		9 &  \qquad $\vec{Z}_{n}=\vec{Z}_{n-1}$.\\
		10 & \textbf{end if} \\
		11 & ${{\vec{E}}_{n}}={{\vec{I}}_{n}}\vec{Z}_{n}+{\vec{V}}_{n}$. \\
		12 & \textbf{for} ${{I}_{d}}={{I}_{d}}+\Delta {{I}_{d}}$.\\
		%${I_{d}}$ is from $I_{dn}$ to $1.3 \sim 1.5 I_{dN}$.\\
		13 & \qquad \qquad $d=1$.\\
    	%14 & \qquad \qquad ${{\left( E_{dr}-E_{dr0} \right)}^{2}}$$+{{\left( E_{di}-E_{di0} \right)}^{2}}=1.  $\\
		14 & \qquad \textbf{while} $d>\mu  $. \\
		15 & \qquad \qquad $E_{dr}=E_{dr0}$, $ E_{di}=E_{di0}$. Calculate $P_{dr}$, $Q_{dr}$, $P_{di}$, $Q_{di}$, $Q_{acr}$ and $Q_{aci}$ by (\ref{e6})-(\ref{e11}) and (\ref{e3}).\\
		16 & \qquad \qquad $ P_{ar}=P_{r}, Q_{ar}=Q_r-Q_{cr}, P_{ai}=P_i, Q_{ai}=Q_{ci}-Q_i.$ Calculate $E_{dr}$ and $E_{di}$ by (\ref{e4}) and (\ref{e5}).\\
		17 & \qquad \qquad  $d={{\left( E_{dr}-E_{dr0} \right)}^{2}}+{{\left( E_{di}-E_{di0} \right)}^{2}}$, $E_{dr0}=E_{dr}$, $ E_{di0}=E_{di}$.\\
		18 & \qquad  \textbf{end while} \\
		19 & \qquad  \textbf{if} $P_{dr} $ or $P_{di}$ is smaller than the value in the last iteration, or the variables of (\ref{constraint}) exceeds the constraints, \textbf{then break}\\
		20 &  \qquad  \textbf{end if}\\
		21 &  \textbf{end for}\\
		22 & ${\textbf{HVDC-MC}}_{n}$= $ P_{dr} $ or $ P_{di} $ \\
		\hline \hline
	\end{tabular}
	\label{HVDC}
\end{table*}

The response rate of the potential under the load is lower than that under the no-load condition \cite{sauer1998power}. Then we have $ \bigl| \vec{d{{E}_{Gi}}} \bigr| \leq d{{{{E}'}_q}_{\max }} $ ,which results in
\begin{equation}
\bigl| \vec{dE} \bigr|\leq\frac{\sum\limits_{i=1}^{m}{ \bigl| \vec{{{dE}_{Gi}}}}  \bigr|\bigl| {{B}_{i}} \bigr|} {\sum\limits_{i=1}^{m}{ {\bigl|{B}_{i}\bigr|}}}\leq\frac{ {{{{dE}'}_q}_{\max}} \sum\limits_{i=1}^{m}{{}} \bigl|{{B}_{i}}\bigr|}{\sum\limits_{i=1}^{m}{ {\bigl|{B}_{i}\bigr|}}}= d{{{{E}'}_q}_{\max }}.
\end{equation}
Based on the above deviation, {{it can be concluded that the variation of the TE potential will not be larger than that of the generator with the greatest voltage change in the system. Therefore, $d{{{{E}'}_q}_{\max }}$ can be used as the upper bound of $\vec{|dE|}$.}} (Note that if there are other dynamic regulators in the vicinity of the HVDC, we should consider the fastest regulator.)
%$\bigl| \vec{dE} \bigr|$ will not exceed 0.01512 $ p.u. $ within the sampling interval of 10$\ ms$.  %{{%Assuming 0.5 $p.u.$ $\leq \bigl| \vec{E} \bigr|\leq$ 2 $p.u.$,
{{Then, substituting ${\bigl| \vec{dE} \bigr|}_{max}$= 0.01512 $ p.u. $ and ${\bigl| \vec{E} \bigr|}_{min}$= 0.5 $ p.u. $ into (\ref{e36}), yields the threshold
\begin{equation}
\bigl| \vec{dI} \bigr|> 0.03024 \bigl| {\vec{I}} \bigr|.
\label{e44}
\end{equation}}}
In order to obtain smaller estimation error, the threshold in (\ref{e44}) can be appropriately increased. However, if the threshold is too large, it can lead to low measurement redundancy. The simulation results show that the threshold $0.1\bigl| {\vec{I}} \bigr| \sim 0.2\bigl| {\vec{I}} \bigr|$ works well for different operating conditions.

Considering that the change in the potential becomes smaller in the consequent period, the threshold can be reduced appropriately so as to enlarge the sampling interval for more measurement redundancy. This enables us to better handle the measurement error and model uncertainties. In this paper, the following adaptive threshold function is developed:
  \begin{equation}
 \sigma =0.1\sim 0.2\bigl| {{{\vec{I}}}_{n}} \bigr|\left( 1-{{\lambda }^{n-{{n}_{0}}}} \right),
 \label{e43}
 \end{equation}
where $\lambda <1$, $n$ is the sampling time, ${{n}_{0}}$ is the initial triggering time. When $\bigl| d{{{\vec{I}}}_{n}} \bigr|>0.1\sim0.2\bigl| {{{\vec{I}}}_{n}} \bigr|$, the threshold function is used to select the qualified measurements. The closer $\lambda $ is to 1, the slower the function transition is. $\lambda $ can be adjusted according to the HVDC response speed.

\vspace{-0.3cm} 
{{\subsection{Algorithm Implementation}
The detailed steps to implement the proposed HVDC-MC estimation algorithm are displayed in Table \ref{HVDC}. In step 1, the voltage and current measurements from PMU are obtained, followed by the variation calculations of voltage and currents. Steps 3-5 implement the proposed screening strategy for the current measurements. From steps 6-11, the TE impedance is estimated by TLS, followed by the calculation of TE potential. In step 6, $ \vec{E} $, $X$ and $R$ can be solved by (\ref{ee19}) at time $n-1$ and $n$. The limitations can be adjusted based on the actual condition. From steps 12-22, the power flow algorithm for AC-DC hybrid systems is performed and the HVDC-MC is estimated considering all constraints.}}

\section{Numerical Results}
{{A tutorial example on a two-bus system connected by an HVDC will be first used to discuss the constraints of HVDC-MC. 
%Then, larger-scale IEEE 39-bus system is utilized to test the scalability and effectiveness of the proposed method
Then, the larger-scale IEEE 39-bus system is utilized to evaluate the estimation of TE and HVDC-MC.
An application for multi-DC coordinated control is also presented based on HVDC-MC estimation results as a use case. All the simulations are carried out on the software PSD-BPA. The HVDC control model is based on the CIGRE HVDC benchmark system \cite{szechtman1991benchmark}. The system frequency is 50 Hz and the base MVA is 1000 $MVA$;
the AC voltage base	is 345 $kV$; the nominal DC voltage and current are	500 $kV$ and 2 $kA$, respectively; ${\alpha}_{min}=5^{\circ }$, $ {\gamma}_{min} =17^{\circ }$, $E_{min}$= 0.9 $p.u.$, $B$ = 2, ${{R}_{d}}$ = $5.79$ $\Omega $, $X_{dr}$ = $8.3201$ $\Omega $, $X_{di}$ = $7.1949$ $\Omega $. For VDCOL, $V_{1}$ = 0.4 $p.u.$, $V_{2}$ = 0.9 $p.u.$, $I_{1}$ = 0.55 $p.u.$, $I_{2}$ = 1 $p.u.$, $k_{1}$ = 0.9, $k_{2}$ = 1, $V_{d}$ is the DC voltage at the middle of the DC line. This paper assumes that HVDC has a long term overloading capacity of 1.3 times. In the steady-state condition, the HVDC rectifier will adjust the initial ignition angle to $15^\circ$ through the transformer tap.
%{{\emph{Remark 1}: in the simulation, the rated power of HVDC is 1000 $MW$ and the initial power of HVDC in the examples is set lower than the nominal value. This is reasonable because that HVDC does not always run at the nominal power as it depends on the power demand. Moreover, for some weak system, HVDC may also not run at the nominal power due to the N-1 or some other security stability constraint.% Monitoring the real-time available power is helpful for operators to master the operation and controllable state of HVDC.
%Furthermore, in the case of faults and if the AC-line is disconnected, HVDC is likely to be operated under non-nominal power operation. In this paper, the HVDC maximum emergency power can be directly estimated based on local measurements when the model and constraints are determined.
\vspace{-0.3cm} 
\subsection{Case 1: Two-Bus System Connected by HVDC}
We first carry out simulations on the system shown in Fig. \ref{1} to investigate the HVDC-MC under different constraints. Here, $E_r$ = $E_i$ = 1 $p.u.$ and $Q_{acr}$ = $Q_{aci}$ = 300 $Mvar$ at the rated voltage. The initial HVDC power is 600 $MW$. Table \ref{ta3} displays the estimation results of HVDC-MCs at different system strengths, i.e., 861.8 $MW$, 958.3 $MW$ and 758.8 $MW$ that are limited by ${\alpha}_{min}$, ${I}_{VD}$ and ${E}_{min}$, respectively. Compared with the simulation results, the errors are 0.47 \% , 0.17 \% and 0.04 \%, respectively.
\vspace{-0.3cm} 
\begin{table}[H]
	\centering
	\caption{{{HVDC-MC under different system strengths}} }
	\begin{tabular}{c c c c}
	\hline \hline
Parameter  &	Example 1 & Example	2& Example 3 \\
	\hline 
$X_{r}$ &	0.2 $p.u.$ &  0.1 $p.u.$ & 0.1 $p.u.$\\
$X_{i}$	&  0.01 $p.u.$ &  0.2 $p.u.$ & 0.4 $p.u.$\\
$N_{r}$ &  0.5738 &  0.5732  & 0.5738 \\
$N_{i}$	&  0.5718 &  0.5718  & 0.5765 \\
Constraint & ${\alpha}_{min}$ & ${I}_{VD}$ & ${E}_{min}$ \\
Proposed method & 861.8 $MW$ & 937.9 $MW$ & 758.8 $MW$ \\
Simulation & 857.8 $MW$ & 936.3 $MW$ & 756.5 $MW$ \\
Error      & 0.47 \% & 0.17 \% & 0.04 \% \\
    \hline \hline
	\end{tabular}
	\label{ta3}
\end{table}
%\vspace{-0.3cm} 
Fig. \ref{51} shows the curves of the ignition angle, extinction angle, AC/DC voltage and DC current with the increase of DC power in Example 1. It can be seen from Fig. \ref{51}(b), (c) and (d) that the ignition angle reaches the constraint during the power growth while there are margins for $ I_{VD} $, $ I_{RA} $ and $ E_{min} $.
\begin{figure}[H]
\setlength{\abovecaptionskip}{-0.1cm}   %调整图片标题与图距离
	\centering
	\includegraphics[scale=0.7] {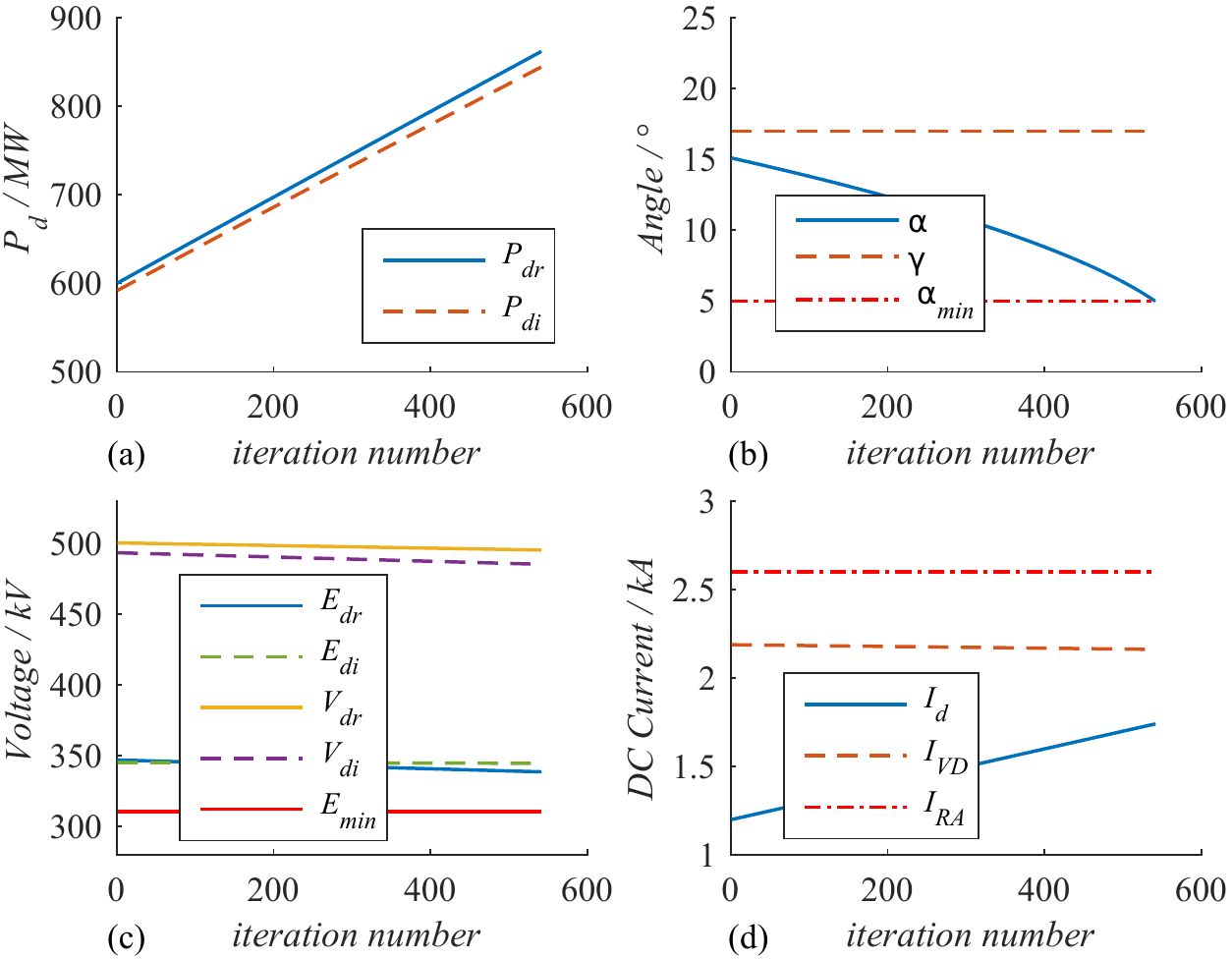}
	\caption{{{HVDC-MC using proposed method. (a) Active power of HVDC; (b) Ignition angle and extinction angle; (c) AC/DC voltage magnitudes; (d) DC current.}}}
	\label{51}
\end{figure}
%\vspace{-0.3cm} 
Fig. \ref{52} shows the dynamic curves of the DC power, regulation angle, AC/DC voltage and DC current from the simulations. It can be found that HVDC increases the power to 857.8 $MW$ in 1 s. The comparison results of the proposed method with the simulation results are demonstrated in Table \ref{ta4}, where the error is between $-0.18 \%$ and $0.47 \%$. In this case, the algorithm can accurately calculate the HVDC-MC under various constraints with the known system parameters. Next, we will use the local measurements to estimate the TE parameters and HVDC-MC.
\vspace{-0.2cm} 
\begin{figure}[H]
\setlength{\abovecaptionskip}{-0.1cm}   %调整图片标题与图距离
	\centering
	\includegraphics[scale=0.7] {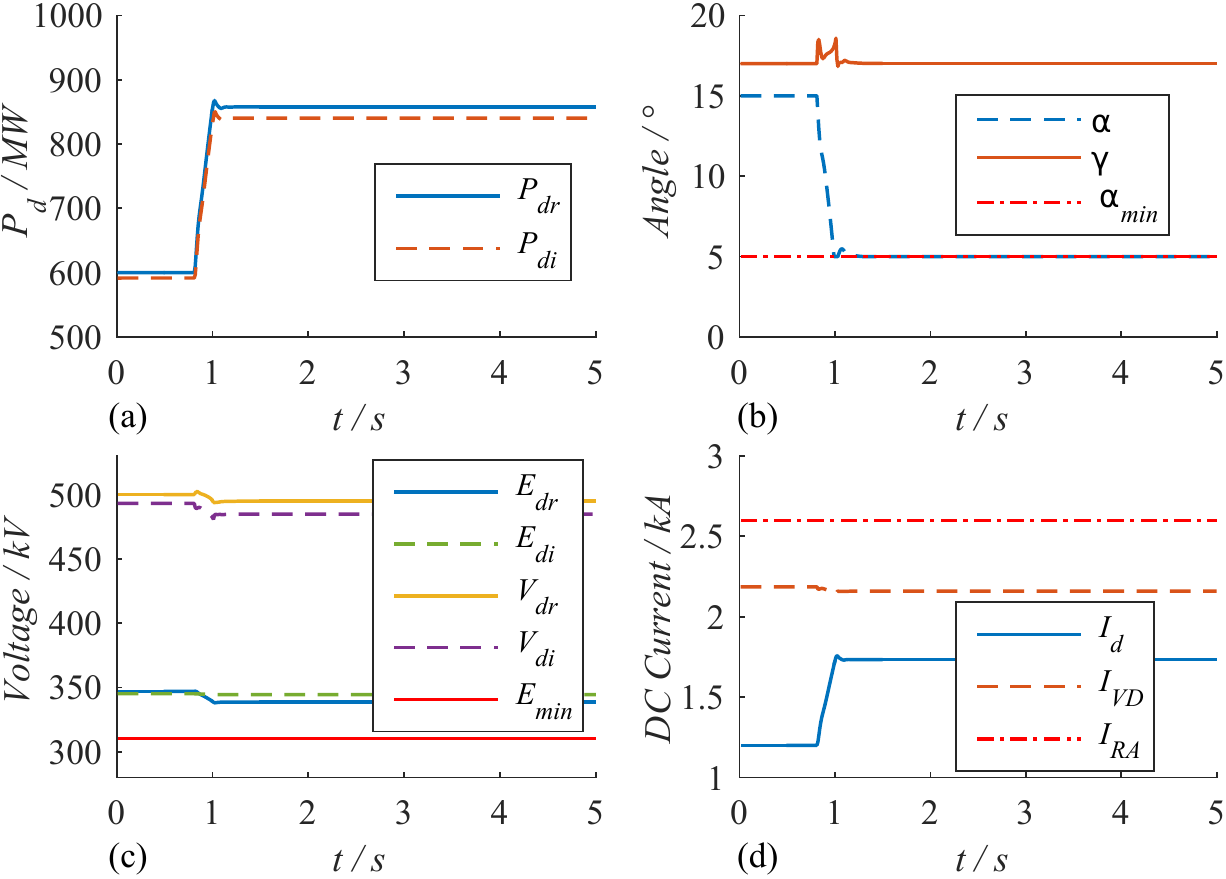}
	\caption{{{HVDC-MC with constraint ${\alpha}_{min}$. (a) Active power of HVDC; (b) Ignition angle and extinction angle; (c) AC/DC voltage magnitudes; (d) DC current.}}}
	\label{52}
\end{figure}
\vspace{-0.3cm}
\begin{table}[H]
	\centering
	\caption{{{Comparison between the proposed method and simulation}}}
	\begin{tabular}{c c c c}
	\hline \hline
Parameter &	Proposed method &	Simulation & Error \\
	\hline 
$P_{dr}$ &	861.8 $MW$ &  857.8 $MW$ &0.47\%  \\
$P_{di}$ &  844.3 $MW$ &  840.8 $MW$ & 0.42\% \\
$\alpha$ &  $5.005^{\circ}$ &  $5^{\circ}$  & 0.10\% \\
$\gamma$ &  $17^{\circ}$   &   $17.03^{\circ}$  & -0.18\% \\
$E_{dr}$ & 338.7 $kV$ & 338.7 $kV$ & 0.00\% \\
$E_{di}$ & 344.7 $kV$ & 344.5 $kV$ & 0.06\% \\
$V_{dr}$ & 495.3 $kV$ & 494.8 $kV$ & 0.10\% \\
$V_{di}$ & 485.2 $kV$ & 484.8 $kV$ & 0.08\% \\
$I_{d}$  & 1.74  $kA$ & 1.734 $kA$ & 0.35\% \\
    \hline \hline
	\end{tabular}
	\label{ta4}
\end{table}
%\vspace{-0.3cm}
%\vspace{-0.5cm}
\begin{figure}[H]
\setlength{\abovecaptionskip}{-0.1cm} 
	\centering
	\includegraphics[scale=0.4] {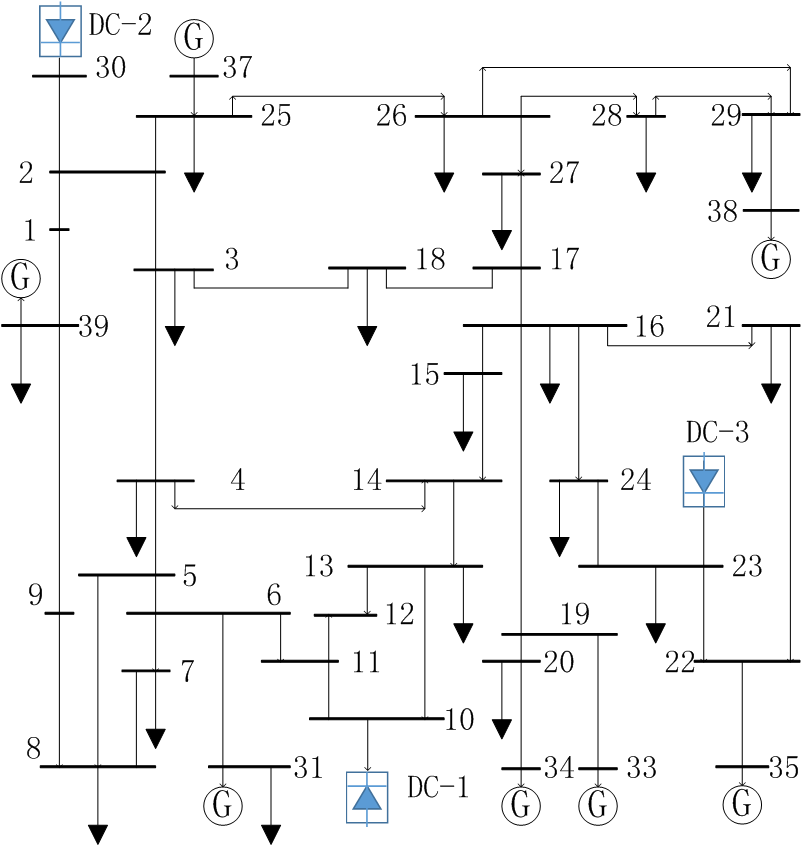}
	\caption{One-line diagram of IEEE-39 system with three HVDCs}
	\label{5}
\end{figure}
%\vspace{-0.1cm}
\subsection{Case 2: IEEE 39-bus System with HVDCs}
Three HVDCs are added to the IEEE-39 system. They are connected to Bus 10, Bus 30 and Bus 23, respectively, and the schematic of the system is displayed in Fig. \ref{5}. The sub-transient model with the excitation system is assumed for all the generators \cite{gomez2011prediction}. The kinetic energy of generator 39 is changed to 5000 $MW$$\cdot$$s$. The frequency coefficient of the governor is 0.05 and the dead-band is 0.2 Hz. The HVDC parameters are shown in Table \ref{ta5}.}}
\vspace{-0.3cm} 
\begin{table}[H]
	\centering
	\caption{{{The system data}} }
	\begin{tabular}{c c c c}
	\hline \hline
 Parameter &	DC 1 & DC 2& DC 3 \\
	\hline 
$P_{dr}$ &	600 $MW$ &  400 $MW$ & 500 $MW$\\
$E_{r}$ &	1 $p.u.$ &  1 $p.u.$ & 1 $p.u.$\\
$X_{r}$ &	0.1 $p.u.$ &  0.3 $p.u.$ & 0.1 $p.u.$\\
%$X_{i}$	&  0.01 $p.u.$ &  0.2 $p.u.$ & 0.4 $p.u.$\\
$N_{r}$ &  0.5767 &  0.5668  & 0.5784 \\
$N_{i}$	&  0.5596 &  0.5423  & 0.5540 \\
$Q_{acr}$ &  260 $Mvar$ &  200 $Mvar$  & 200 $Mvar$ \\
$Q_{aci}$ &   300 $Mvar$ &  200 $Mvar$  & 200 $Mvar$ \\
    \hline \hline
    \multicolumn{4}{r}{$Q_{ac}$ is the capacity at rated voltage.} \\
	\end{tabular}
	\label{ta5}
\end{table}
\vspace{-0.3cm} 
\subsubsection{HVDC-MC Estimation Neglecting AC System Dynamics}
%First of all, benchmark TE impedance is calculated. This can be done by setting all generators with the infinite inertia and neglecting the excitation system. Then, we increase the DC power continuously and the TE impedance can be calculated by using any two different operating points, yielding the true value 0.235$\ p.u.$. Then 
we first use the classic model for generators with the constant voltage behind transient reactance and the infinity inertia. Increasing the HVDC power continuously, the TE impedance can be calculated by using any two different operating points, yielding the true value 0.235$\ p.u.$. Then we further consider the disturbance case. A three-phase short-circuit fault is applied to the AC line 10-13 at 0.2 $s$, and the fault is cleared at 0.28 $ s $. {{HVDC starts to recover power at 0.53 $s$. The curves of the threshold function (\ref{e43}) and $\bigl| \vec{dI}_{n} \bigr| $ are shown in Fig. \ref{TES}(a). The measurements that are larger than the threshold are selected. Then, the equivalent reactance is identified as 0.2317$\ p.u.$ at 0.55 $s$ as shown in Fig. \ref{TES}(b).
%\vspace{-0.5cm}
\vspace{-0.2cm} 
\begin{figure}[H]
\setlength{\abovecaptionskip}{-0.5cm}   %调整图片标题与图距离
	\centering
	\includegraphics[scale=0.7] {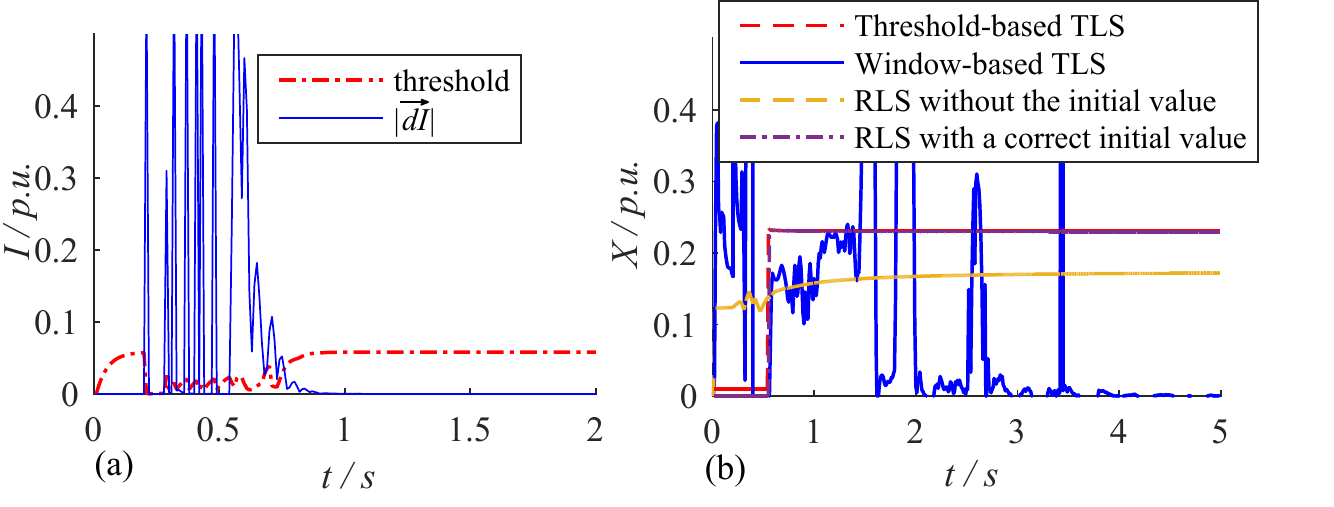}
	\caption{{{TE estimation neglecting AC system dynamics. (a) Threshold function screening; (b) TE reactance estimation.}}}
	\label{TES}
\end{figure}
\vspace{-0.3cm}
{{TLS and the recursive least square (RLS) are common methods of tracking parameter changes. The window-based TLS method uses a moving measurements window to estimate the parameters, which requires that the parameters in the window are constant. Moreover, the measurements in the window should ensure the observability. In this case, TE parameters are constant after the fault by neglecting AC system dynamics. However, we can see from Fig. \ref{TES}(a) that there is only a transient current change. After that, the measurements are invariant, which leads to the scenario that the parameters are not observable. Therefore, the window-based TLS leads to the incorrect estimation results. The RLS method requires an appropriate initial value and the parameters for estimation should also be constants. Fig. \ref{TES}(b) shows that RLS cannot accurately estimate the parameters if the provided initial value is not correct.}}

Based on the estimated TE parameters, Fig. \ref{MCpaper} (a) shows the real-time HVDC-MC is 922.9 $MW$. Figs. \ref{MCpaper} (b), \ref{MCpaper} (c) and \ref{MCpaper} (d) are respectively the ignition angle, the voltage and the DC current corresponding to the maximum power. We can observe that the VDCOL limits the DC current $I_{d}$. 
\vspace{-0.3cm}
\begin{figure}[H]
\setlength{\abovecaptionskip}{-0.1cm}   
	\centering
	\includegraphics[scale=0.7] {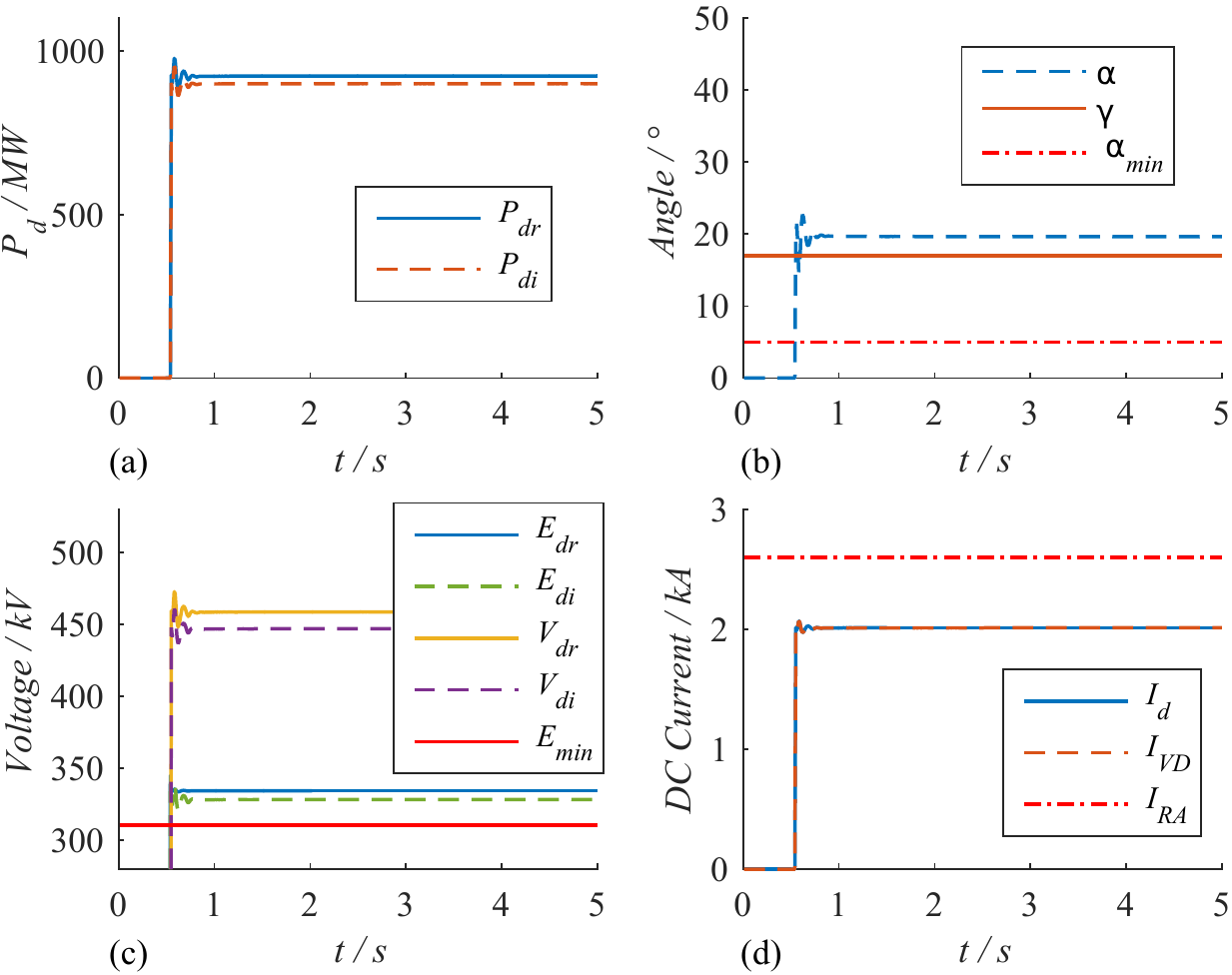}
	\caption{{{HVDC-MC neglecting AC system dynamics using proposed method. (a) Active power of HVDC; (b) Ignition angle and extinction angle; (c) AC/DC voltage magnitudes; (d) DC current.}}}
	\label{MCpaper}
\end{figure}
\vspace{-0.3cm} 
To verify above results, the power order of 1000 MW is provided to the HVDC in 1 s. Fig. \ref{MCsimu}(a) shows the simulated HVDC power and it can be found that the power of HVDC only increases to 954.8 $MW$ in 1.2 $s$. From Fig. \ref{MCsimu}(d), it can be observed that due to the DC current limitation by VDCOL, the DC current can no longer increase. Fig. \ref{MCsimu}(b) and Fig. \ref{MCsimu}(c) display the ignition/extinction angle and AC/DC voltage magnitudes, which do not reach the limits. 
\vspace{-0.3cm} 
\begin{figure}[H]
\setlength{\abovecaptionskip}{-0.1cm}   %调整图片标题与图距离
	\centering
	\includegraphics[scale=0.7] {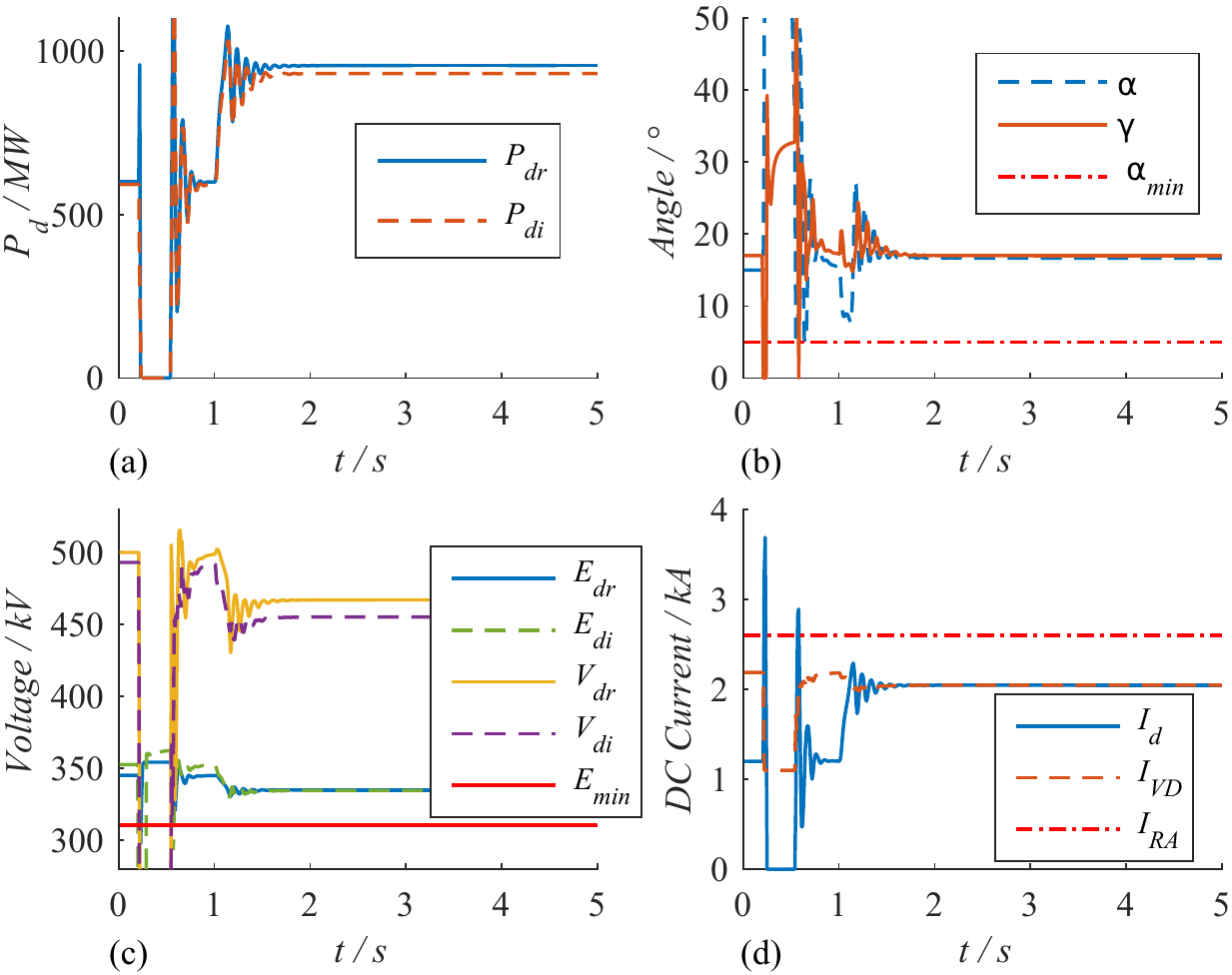}
	\caption{{{HVDC-MC neglecting AC system dynamics from the simulation. (a) Active power of HVDC; (b) Ignition angle and extinction angle; (c) AC/DC voltage magnitudes; (d) DC current.}}}
	\label{MCsimu}
\end{figure}
\vspace{-0.3cm} 
Table \ref{ta9} shows the detailed comparison results between the proposed method and the simulations. Except for the large variations of the ignition angle, other variables are in a reasonable range. 
\vspace{-0.3cm}

\begin{table}[H]
	\centering
	\caption{{{Comparison between the proposed method and simulation}}}
	\begin{tabular}{c c c c}
	\hline \hline
Parameter &	Proposed method &	Simulation & Error \\
	\hline 
$P_{dr}$ &	922.9 $MW$ &  954.9 $MW$ & -3.35\%  \\
$P_{di}$ &  899.4 $MW$ &  930.7 $MW$ & -3.36\% \\
$\alpha$ &  $19.65^{\circ }$ &  $16.66^{\circ }$  & 17.95\% \\
$\gamma$ &  $17^{\circ }$   &   $17^{\circ }$  & 0.00\% \\
$E_{dr}$ & 334.4 $kV$ & 335 $kV$ & -0.18\% \\
$E_{di}$ & 328.3 $kV$ & 334.6 $kV$ & -1.88\% \\
$V_{dr}$ & 458.6 $kV$ & 467 $kV$ & -1.80\% \\
$V_{di}$ & 446.9 $kV$ & 455.2 $kV$ & -1.82\% \\
$I_{d}$ &  2.011 $kA$ &   2.045 $kA$ & -1.66\% \\
    \hline \hline
	\end{tabular}
	\label{ta9}
\end{table}
%\vspace{-0.3cm} 

\subsubsection{HVDC-MC Estimation Considering AC System Dynamics}
the impacts of AC system dynamics are considered for HVDC-MC estimation. The fault condition is the same as previous section and the curves of the threshold function and $\bigl| \vec{dI}_{n} \bigr| $ are displayed in Fig. \ref{TEd_TLS}(a). The equivalent reactance is identified as 0.2221$\ p.u.$ at 0.55 s as shown in Fig. \ref{TEd_TLS}(b). {{In this case, the TE potential after the fault is not constant due to AC system dynamics. Fig. \ref{TEd_TLS}(b) shows that RLS cannot accurately estimate the parameters even there is a correct initial value. This demonstrates the benefit of the adaptive measurement selection for the TE parameters estimation. Fig. \ref{different_threshold} shows the TE estimation when different thresholds are used. It can be found that when the threshold is larger than 0.03024, the estimation result is very close to the true value. With a further increase of  the threshold, the estimation results become more accurate. This justifies the choice of threshold $0.1\bigl| {\vec{I}} \bigr| \sim 0.2\bigl| {\vec{I}} \bigr|$ in this paper.}}
\vspace{-0.3cm} 
\begin{figure}[H]
\setlength{\abovecaptionskip}{-0.1cm}   %调整图片标题与图距离
	\centering
	\includegraphics[scale=0.7] {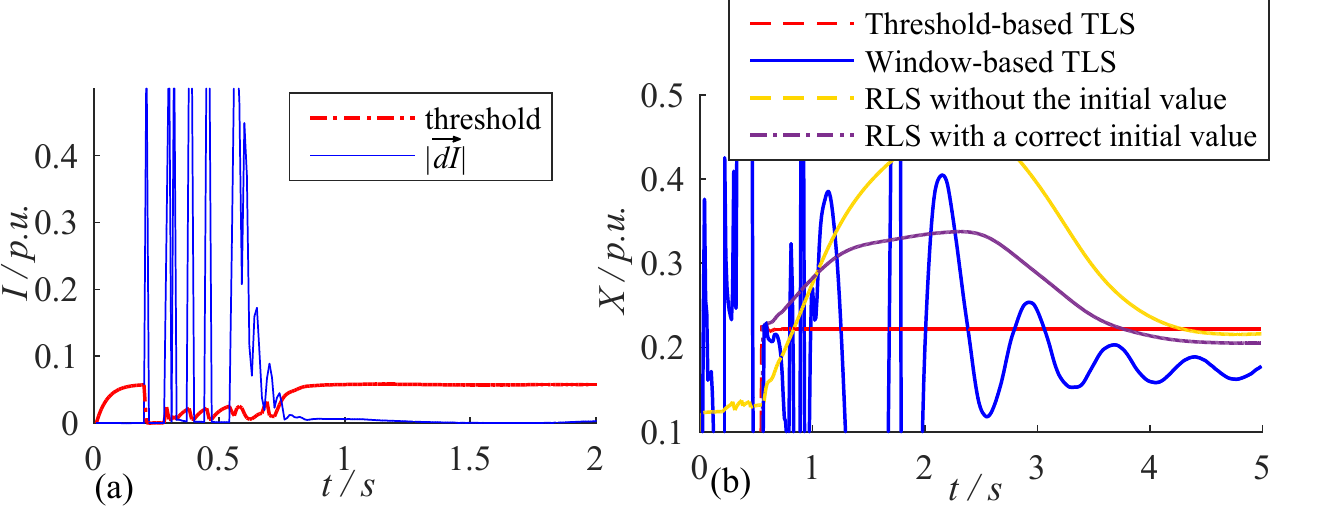}
	\caption{{{TE estimation considering AC system dynamics (a) Threshold function screening; (b) TE reactance estimation.}}}
	\label{TEd_TLS}
\end{figure}
\vspace{-0.3cm} 
\begin{figure}[H]
\setlength{\abovecaptionskip}{-0.1cm}   %调整图片标题与图距离
	\centering
	\includegraphics[scale=0.7] {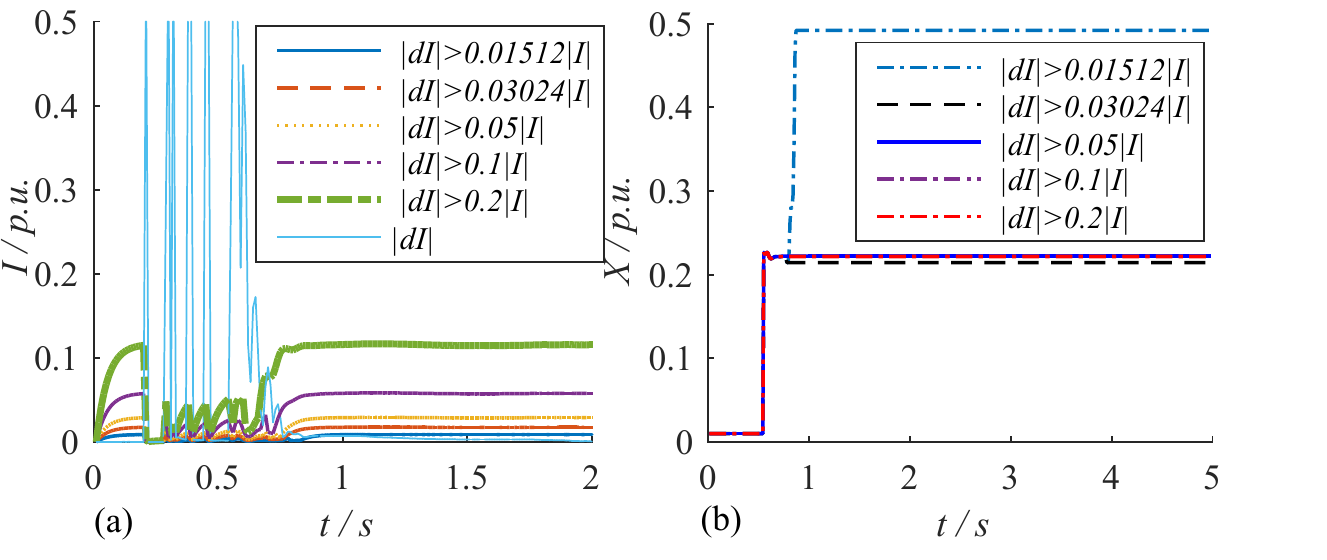}
	\caption{{{TE estimation considering different thresholds (a) Threshold for measurement screening; (b) TE reactance estimation.}}}
	\label{different_threshold}
\end{figure}

The HVDC-MC corresponding to the proposed method and the simulations are shown in Fig. \ref{MCdypaper} and Fig. \ref{MCdysimu}, respectively. Compared to the previous case, the HVDC-MC changes with time since the TE potential changes after the fault. HVDC-MC is about 940 $MW$ calculated using proposed method while the simulation result is 970 MW. Note that the power increase is constricted by VDCOL. The maximum DC currents are about 2.036 $kA$ and 2.07 $kA$, respectively. According to these results, the effectiveness of the proposed method can be demonstrated.}}

\begin{figure}[H]
\setlength{\abovecaptionskip}{-0.1cm}   
	\centering
	\includegraphics[scale=0.7] {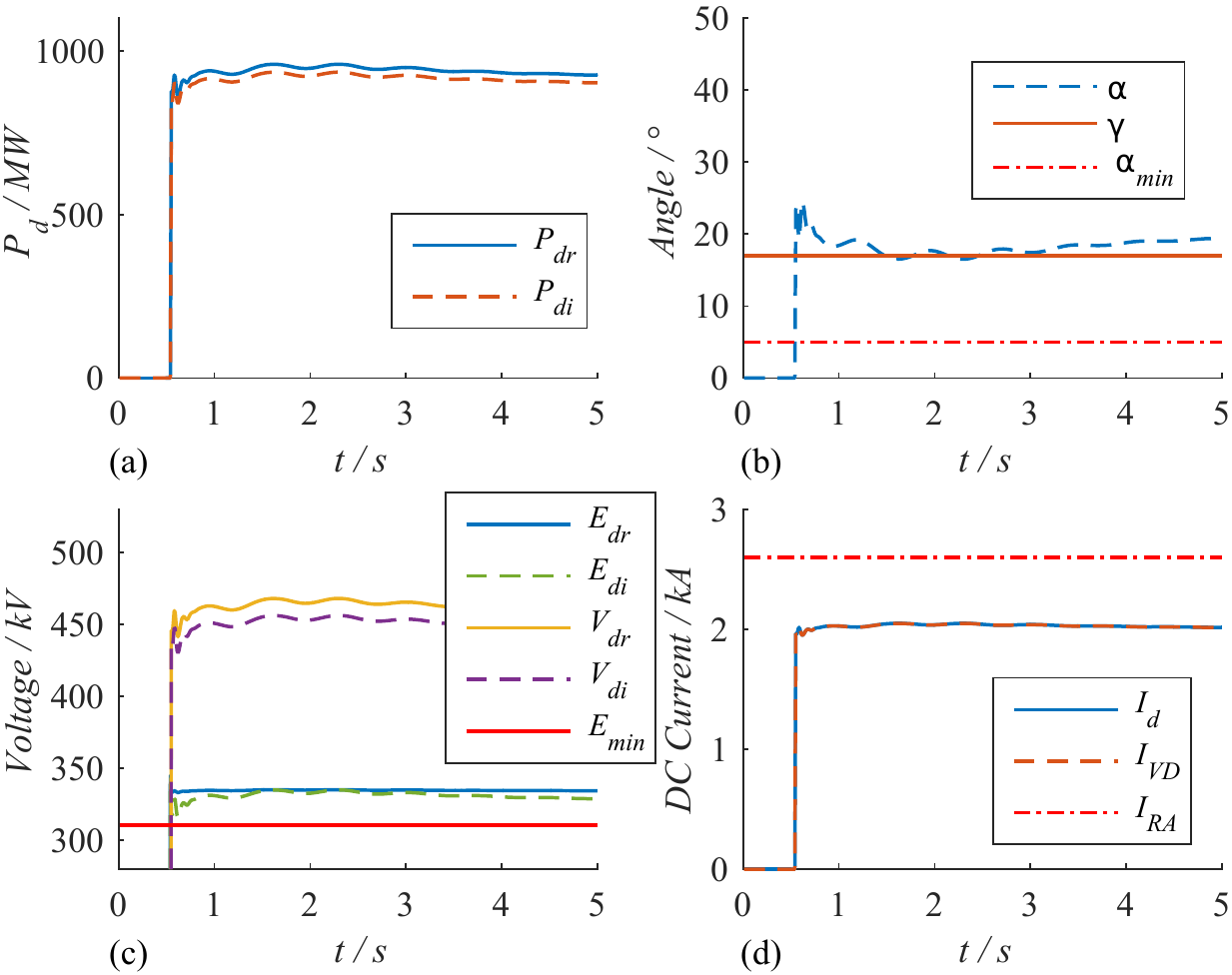}
	\caption{{{HVDC-MC considering AC system dynamics using proposed method. (a) Active power of HVDC; (b) Ignition angle and extinction angle; (c) AC/DC voltage magnitudes; (d) DC current.}}}
	\label{MCdypaper}
\end{figure}
%\vspace{-0.3cm} 

\begin{figure}[H]
\setlength{\abovecaptionskip}{-0.1cm}   %调整图片标题与图距离
	\centering
	\includegraphics[scale=0.7] {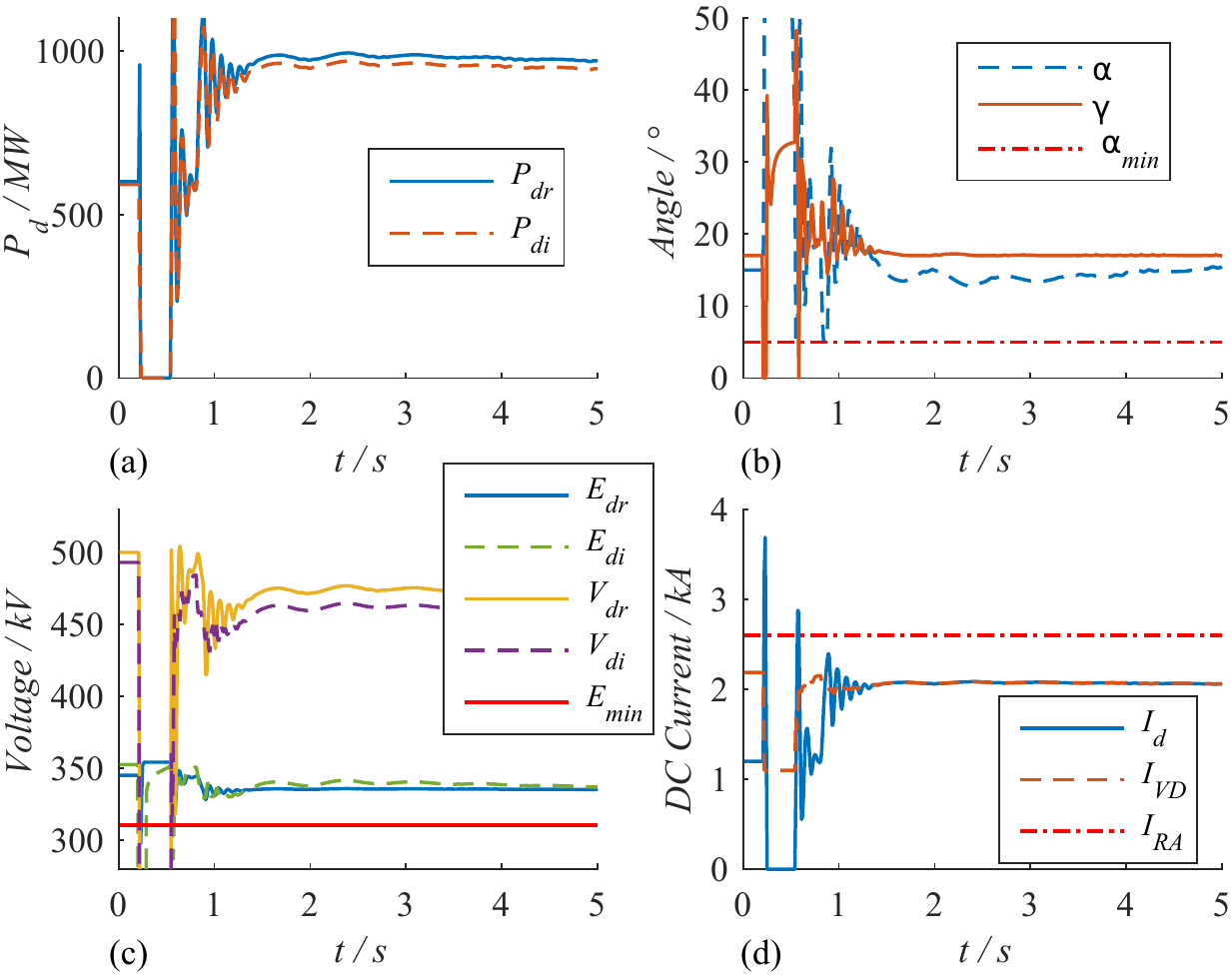}
	\caption{{{HVDC-MC considering AC system dynamics from the simulation. (a) Active power of HVDC; (b) Ignition angle and extinction angle; (c) AC/DC voltage magnitudes; (d) DC current.}}}
	\label{MCdysimu}
\end{figure}
%\vspace{-0.3cm} 

\vspace{-0.3cm}
\subsection{Case 3: HVDC-MC Estimation Considering Measurement Noise/Error}
To further show the capability of the proposed method in dealing with measurement noise/error, a Gaussian noise with zero-mean and variance of ${10}^{-3}$ is added to the PMU measurements. Fig. \ref{noise}(a) displays the reactance estimation error under 1000 sets of random noise. {{Under the worst scenario, that is $1.50\%$ estimation error, the impedance is 0.2254 $ p.u.$. Using this impedance, HVDC-MCs calculated with and without noise are shown in Fig. \ref{noise}(b). It can be concluded that the measurement noise has limited influence on HVDC-MC estimation.

\begin{figure}[H]
\setlength{\abovecaptionskip}{-0.1cm}   %调整图片标题与图距离
	\centering
	\includegraphics[scale=0.68] {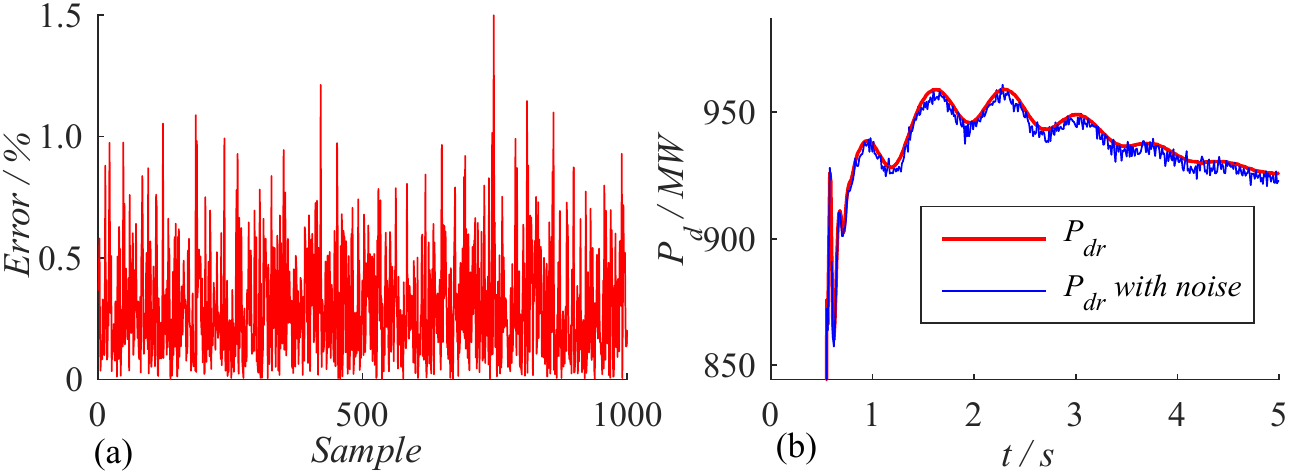}
	\caption{{{Estimation error considering measurement noise. (a) Reactance estimation error (b) HVDC-MC estimation with and without noise.}}}
	\label{noise}
\end{figure}

\vspace{-0.3cm} 
\subsection{Case 4: Multi-DC Coordinated Control Based on HVDC-MC Estimation}
In the case of cascading scenarios, the benefit of the proposed method for multi-DC coordinated control is analyzed. It should be noted that multi-infeed DC is a typical application scenario of the proposed method as the practical systems may have multi-HVDC \cite{shao2017fast}. The available capacity of each HVDC will be identified separately and this allows us to allocate the required control power appropriately.
Here, the HVDC model is based on the ABB actual control system and the DC current of VDCOL is unlimited when the DC voltage on both sides is higher than 0.8p.u..}} It is assumed that line 14-15 and line 13-14 are tripped at 0.2 $ s $ and 0.24 $ s $, respectively. In addition, at 0.4 $ s $, both DC-1 and DC-3 have commutation failure with a duration of 200 $ ms $. After that, DC-2 is blocked at 0.5 $ s $. The reactive power compensation of DC-2 is cut off at 0.7 $ s $. Using the proposed method, the TE impedance of DC-1 and DC-3 are estimated to be 0.3266 $ p.u.$ and 0.2464 $p.u.$ at 0.6 $ s $, respectively. Fig. \ref{11} (a) and (b) display the magnitudes of two DC real-time TE potentials and the HVDC-MC, respectively.
%%\vspace{-0.3cm}
\begin{figure}[htb]
\setlength{\abovecaptionskip}{-0.1cm} 
	\centering
	\includegraphics[scale=0.65] {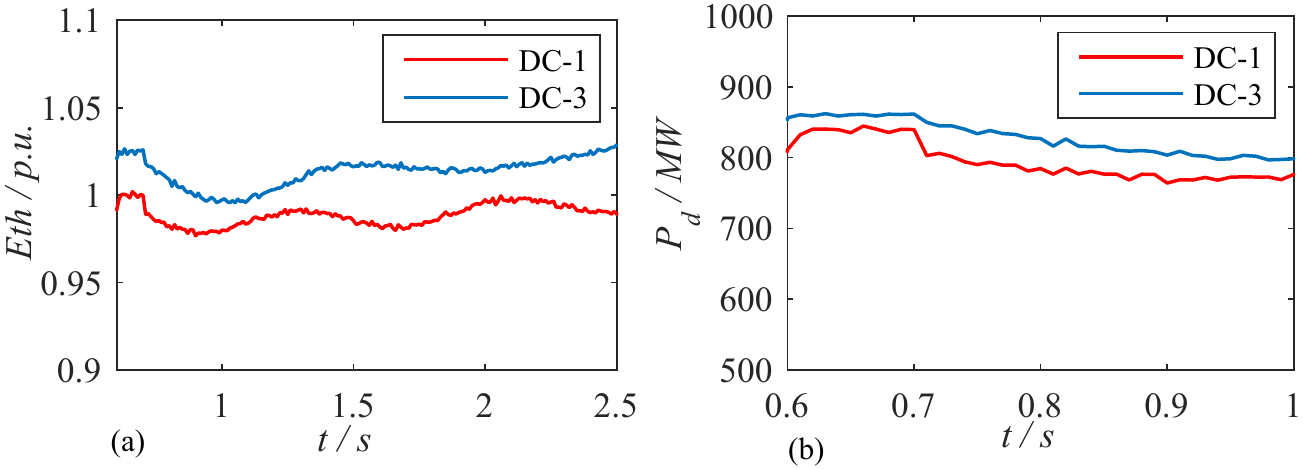}
	\caption{HVDC-MC for multi-DC coordinated control. (a) the real-time TE potential magnitude; (b) HVDC-MC for DC-1 and DC-3.}
	\label{11}
\end{figure} 

At 0.6 $ s $, the HVDC-MCs of DC-1 and DC-3 are 810 $ MW $ and 856 $ MW $, respectively. The constraints are the low AC voltage of the receiving bus of DC-1 and the sending bus of DC-3. Based on the initial operational values, the DC power margins that can be used for emergency support are 210 $ MW $ and 356 $ MW $ while the power shortage is 400 $ MW $. To balance the generation and power, the power of DC-1 and DC-3 should be increased to 727 $ MW $ and 773 $ MW $ with the same remaining margin of 83 $ MW $, respectively. In the follow-up process, real-time monitoring of HVDC-MC is also needed for control adjustment. As shown in Fig. \ref{11}, at 0.7 $ s $, the reactive power compensation of DC-2 is removed, and the TE potential of the system decreases, resulting in the decrease of HVDC-MC.

%By contrast, by using the MC estimation results, appropriate control action can be applied to restore the system frequency.

\begin{figure}[H]
\setlength{\abovecaptionskip}{-0.1cm} 
	\centering
	\includegraphics[scale=0.7] {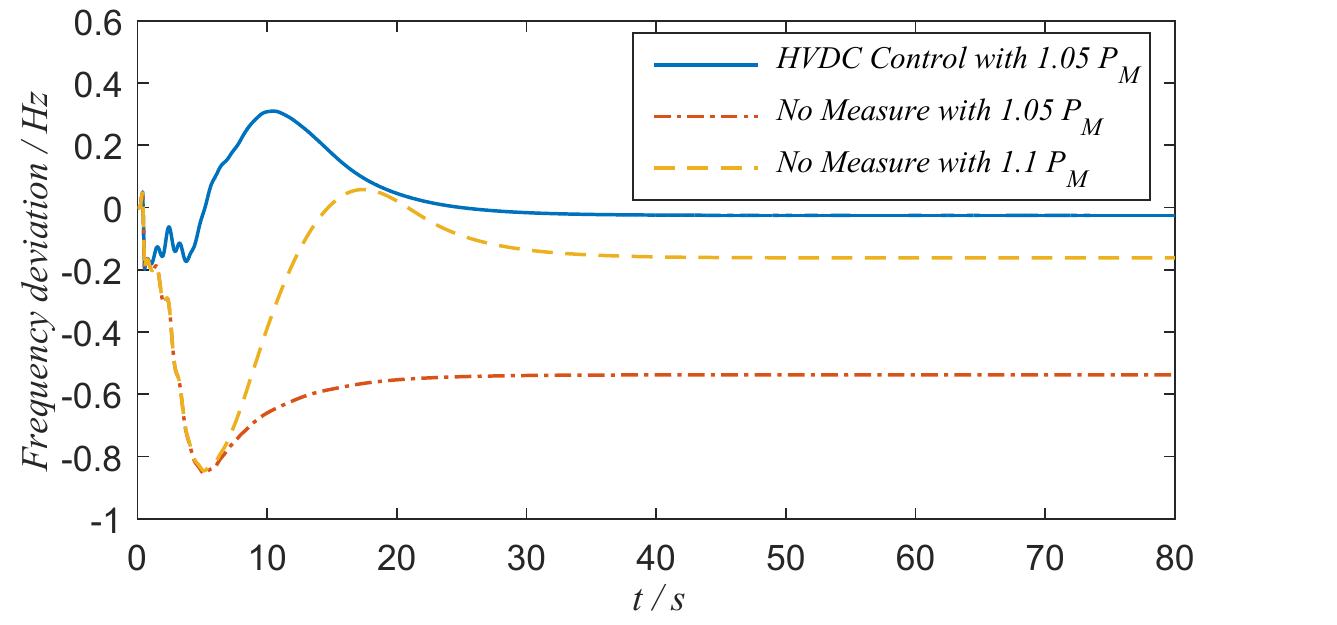}
	\caption{ {{Frequency response with and without the proposed control.}}}
	\label{12}
\end{figure} 
{{Assuming the prime mover has respectively an upper limit of 1.1 times and 1.05 times rated mechanical power $P_M$, the frequency response curves of the system are displayed in Fig. \ref{12}. Each case is analyzed according to the operating standard of instantaneous frequency not exceeding $\pm 0.5$Hz and steady frequency not exceeding $\pm 0.2$ Hz \cite{fang2018inertia}. Note that the off-line planning strategy table does not have the cascading failure countermeasures for this complicated scenario. Consequently, the instantaneous frequency exceeded $-0.5$ Hz in the absence of HVDC emergency control measures. Meanwhile, when the upper limit is 1.05 times $P_M$, the steady-state frequency exceeds $-0.2$ Hz. The steady-state frequency deviation decreases when the upper limit is 1.1 times $P_M$. However, due to the slow response speed of the prime mover, there is little effect on the lowest frequency. Note that the off-line planning strategy is generally not implemented because enforcing a mismatched off-line strategy can cause other risks. If the frequency is out of the range, the local low-frequency load shedding may be triggered. With the proposed method, the estimated HVDC-MC can be effectively leveraged for emergency control, avoiding the load loss.}}
 
{{\emph{Remark}: with the HVDC control, the largest instantaneous over frequency deviation reaches $0.3 $ Hz. Although it does not exceed the instantaneous frequency limitation, there is still room for improvement. For example, real-time monitoring of system frequency and adaptive power regulation can reduce the issue of excessive frequency. However, these need to be done based on the real-time HVDC-MC and the proposed method can help.}}

%\vspace{-0.3cm}
\section{Conclusion and Future Work}
In this paper, a synchrophasor-measurement-based method is developed to assess the maximum power capability the HVDC can instantly provide subject to large disturbances. Unlike the off-line oriented method, our proposed approach uses the online synchrophasor measurements and leads to the online monitoring of HVDC-MC. {{The key idea is to develop an adaptive current measurements selection strategy to enhance the observability of TE parameters. The estimated TE parameters are used together with the VDCOL and AC voltage limitations for HVDC-MC estimation while respecting all regulation characteristics.}} Numerical results demonstrate that the proposed approach can accurately estimate HVDC-MC and assist multi-DC coordinated control. Our future work will be on testing the developed method using field PMU measurements and actual power systems.

\ifCLASSOPTIONcaptionsoff
  \newpage
\fi

\bibliographystyle{IEEEtran}
\bibliography{IEEEabrv,mybibfile}

\begin{thebibliography}{10}
\providecommand{\url}[1]{#1}
\csname url@samestyle\endcsname
\providecommand{\newblock}{\relax}
\providecommand{\bibinfo}[2]{#2}
\providecommand{\BIBentrySTDinterwordspacing}{\spaceskip=0pt\relax}
\providecommand{\BIBentryALTinterwordstretchfactor}{4}
\providecommand{\BIBentryALTinterwordspacing}{\spaceskip=\fontdimen2\font plus
\BIBentryALTinterwordstretchfactor\fontdimen3\font minus
  \fontdimen4\font\relax}
\providecommand{\BIBforeignlanguage}[2]{{%
\expandafter\ifx\csname l@#1\endcsname\relax
\typeout{** WARNING: IEEEtran.bst: No hyphenation pattern has been}%
\typeout{** loaded for the language `#1'. Using the pattern for}%
\typeout{** the default language instead.}%
\else
\language=\csname l@#1\endcsname
\fi
#2}}
\providecommand{\BIBdecl}{\relax}
\BIBdecl

\bibitem{benasla2018hvdc}
M.~Benasla, T.~Allaoui, M.~Brahami, M.~Denai, and V.~K. Sood, ``{HVDC} links
  between {North Africa and Europe}: Impacts and benefits on the dynamic
  performance of the {European} system,'' \emph{Renewable and Sustainable
  Energy Reviews}, vol.~82, pp. 3981--3991, 2018.

\bibitem{alassi2019hvdc}
A.~Alassi, S.~Ba{\~n}ales, O.~Ellabban, G.~Adam, and C.~MacIver, ``{HVDC}
  transmission: technology review, market trends and future outlook,''
  \emph{Renewable and Sustainable Energy Reviews}, vol. 112, pp. 530--554,
  2019.

\bibitem{sun2017renewable}
J.~Sun, M.~Li, Z.~Zhang, T.~Xu, J.~He, H.~Wang, and G.~Li, ``Renewable energy
  transmission by {HVDC} across the continent: system challenges and
  opportunities,'' \emph{CSEE Journal of Power and Energy Systems}, vol.~3,
  no.~4, pp. 353--364, 2017.

\bibitem{shao2017fast}
Y.~Shao and Y.~Tang, ``Fast evaluation of commutation failure risk in
  multi-infeed {HVDC} systems,'' \emph{IEEE Trans. Power Syst.}, vol.~33,
  no.~1, pp. 646--653, 2017.

\bibitem{elizondo2018interarea}
M.~A. Elizondo, R.~Fan, H.~Kirkham, M.~Ghosal, F.~Wilches-Bernal,
  D.~Schoenwald, and J.~Lian, ``Interarea oscillation damping control using
  high-voltage {DC} transmission: A survey,'' \emph{IEEE Trans. Power Syst.},
  vol.~33, no.~6, pp. 6915--6923, 2018.

\bibitem{elizondo2017hvdc}
M.~A. Elizondo, N.~Mohan, J.~O'Brien, Q.~Huang, D.~Orser, W.~Hess, H.~Brown,
  W.~Zhu, D.~Chandrashekhara, Y.~V. Makarov \emph{et~al.}, ``{HVDC} macrogrid
  modeling for power-flow and transient stability studies in north american
  continental-level interconnections,'' \emph{CSEE Journal of Power and Energy
  Systems}, vol.~3, no.~4, pp. 390--398, 2017.

\bibitem{benasla2019power}
M.~Benasla, T.~Allaoui, M.~Brahami, V.~K. Sood, and M.~Dena{\"\i}, ``Power
  system security enhancement by {HVDC} links using a closed-loop emergency
  control,'' \emph{Electric Power Systems Research}, vol. 168, pp. 228--238,
  2019.

\bibitem{liu2017design}
C.~Liu, Y.~Zhao, G.~Li, and U.~D. Annakkage, ``Design of {LCC HVDC} wide-area
  emergency power support control based on adaptive dynamic surface control,''
  \emph{IET Generation, Transmission \& Distribution}, vol.~11, no.~13, pp.
  3236--3245, 2017.

\bibitem{qi2010research}
W.~Qi, Z.~Wenchao, and T.~Yong, ``Research on application of {DC} emergency
  power control to improve the transmission capacity of {UHV} {AC}
  demonstration project,'' in \emph{2010 International Conference on Power
  System Technology}.\hskip 1em plus 0.5em minus 0.4em\relax IEEE, 2010, pp.
  1--5.

\bibitem{gomez2011prediction}
F.~Gomez, ``Prediction and control of transient instability using wide area
  phasor measurements,'' Ph.D. dissertation, PhD thesis, University of
  Manitoba, 2011.

\bibitem{xu2018propagating}
Y.~Xu, L.~Mili, A.~Sandu, M.~R. von Spakovsky, and J.~Zhao, ``Propagating
  uncertainty in power system dynamic simulations using polynomial chaos,''
  \emph{IEEE Trans. Power Syst.}, vol.~34, no.~1, pp. 338--348, 2018.

\bibitem{andersson2005causes}
G.~Andersson, P.~Donalek, R.~Farmer, N.~Hatziargyriou, I.~Kamwa, P.~Kundur,
  N.~Martins, J.~Paserba, P.~Pourbeik, J.~Sanchez-Gasca \emph{et~al.}, ``Causes
  of the 2003 major grid blackouts in {North America and Europe}, and
  recommended means to improve system dynamic performance,'' \emph{IEEE Trans.
  Power Syst.}, vol.~20, no.~4, pp. 1922--1928, 2005.

\bibitem{makarov2012pmu}
Y.~V. Makarov, P.~Du, S.~Lu, T.~B. Nguyen, X.~Guo, J.~Burns, J.~F. Gronquist,
  and M.~Pai, ``{PMU}-based wide-area security assessment: concept, method, and
  implementation,'' \emph{IEEE Trans. Smart Grid}, vol.~3, no.~3, pp.
  1325--1332, 2012.

\bibitem{kamwa2006wide}
I.~Kamwa, J.~Beland, G.~Trudel, R.~Grondin, C.~Lafond, and D.~McNabb,
  ``Wide-area monitoring and control at hydro-qu{\'e}bec: Past, present and
  future,'' in \emph{2006 IEEE Power Engineering Society General
  Meeting}.\hskip 1em plus 0.5em minus 0.4em\relax IEEE, 2006, pp. 12--pp.

\bibitem{zhao2019power}
J.~Zhao, A.~Gomez-Exposito, M.~Netto, L.~Mili, A.~Abur, V.~Terzija, I.~Kamwa,
  B.~C. Pal, A.~K. Singh, J.~Qi \emph{et~al.}, ``Power system dynamic state
  estimation: motivations, definitions, methodologies and future work,''
  \emph{IEEE Trans. Power Syst.}, 2019.

\bibitem{krishayya1997ieee}
P.~Krishayya, R.~Adapa, M.~Holm \emph{et~al.}, ``{IEEE guide for planning DC
  links terminating at AC locations having low short-circuit capacities, part
  I: AC/DC system interaction phenomena},'' \emph{CIGRE, France}, 1997.

\bibitem{balu1994power}
C.~Balu and D.~Maratukulam, \emph{Power system voltage stability}.\hskip 1em
  plus 0.5em minus 0.4em\relax McGraw-Hill, 1994.

\bibitem{yun2019online}
Z.~Yun, X.~Cui, and K.~Ma, ``Online thevenin equivalent parameter
  identification method of large power grids using {LU} factorization,''
  \emph{IEEE Trans. Power Syst.}, vol.~34, no.~6, pp. 4464--4475, 2019.

\bibitem{vu1999use}
K.~Vu, M.~M. Begovic, D.~Novosel, and M.~M. Saha, ``Use of local measurements
  to estimate voltage-stability margin,'' \emph{IEEE Trans. Power Syst.},
  vol.~14, no.~3, pp. 1029--1035, 1999.

\bibitem{wang2012real}
X.~Wang, H.~Sun, B.~Zhang, W.~Wu, and Q.~Guo, ``Real-time local voltage
  stability monitoring based on {PMU} and recursive least square method with
  variable forgetting factors,'' in \emph{IEEE PES Innovative Smart Grid
  Technologies}.\hskip 1em plus 0.5em minus 0.4em\relax IEEE, 2012, pp. 1--5.

\bibitem{chen2014pmu}
C.~Chen, J.~Wang, Z.~Li, H.~Sun, and Z.~Wang, ``{PMU} uncertainty
  quantification in voltage stability analysis,'' \emph{IEEE Trans. Power
  Syst.}, vol.~30, no.~4, pp. 2196--2197, 2014.

\bibitem{babazadeh2017real}
D.~Babazadeh, A.~Muthukrishnan, P.~Mitra, T.~Larsson, and L.~Nordstr{\"o}m,
  ``Real-time estimation of {AC}-grid short circuit capacity for {HVDC} control
  application,'' \emph{IET Generation, Transmission \& Distribution}, vol.~11,
  no.~4, pp. 838--846, 2017.

\bibitem{zhao2016robust}
J.~Zhao, Z.~Wang, C.~Chen, and G.~Zhang, ``Robust voltage instability
  predictor,'' \emph{IEEE Trans. Power Syst.}, vol.~32, no.~2, pp. 1578--1579,
  2016.

\bibitem{su2018robust}
H.-Y. Su and T.-Y. Liu, ``Robust thevenin equivalent parameter estimation for
  voltage stability assessment,'' \emph{IEEE Trans. Power Syst.}, vol.~33,
  no.~4, pp. 4637--4639, 2018.

\bibitem{abdelkader2014online}
S.~M. Abdelkader and D.~J. Morrow, ``Online {Th{\'e}venin} equivalent
  determination considering system side changes and measurement errors,''
  \emph{IEEE Trans. Power Syst.}, vol.~30, no.~5, pp. 2716--2725, 2014.

\bibitem{kundur1994power}
P.~Kundur, N.~J. Balu, and M.~G. Lauby, \emph{Power system stability and
  control}.\hskip 1em plus 0.5em minus 0.4em\relax McGraw-hill New York, 1994,
  vol.~7.

\bibitem{szechtman1991benchmark}
M.~Szechtman, T.~Wess, and C.~Thio, ``A benchmark model for {HVDC} system
  studies,'' in \emph{International conference on AC and DC power
  transmission}.\hskip 1em plus 0.5em minus 0.4em\relax IET, 1991, pp.
  374--378.

\bibitem{lavenius2015performance}
J.~Lavenius, L.~Vanfretti, and G.~N. Taranto, ``Performance assessment of
  {PMU}-based estimation methods of {Thevenin} equivalents for real-time
  voltage stability monitoring,'' in \emph{2015 IEEE 15th International
  Conference on Environment and Electrical Engineering (EEEIC)}.\hskip 1em plus
  0.5em minus 0.4em\relax IEEE, 2015, pp. 1977--1982.

\bibitem{sauer1998power}
P.~W. Sauer and M.~A. Pai, \emph{Power system dynamics and stability}.\hskip
  1em plus 0.5em minus 0.4em\relax Prentice hall Upper Saddle River, NJ, 1998,
  vol. 101.

\bibitem{fang2018inertia}
J.~Fang, H.~Li, Y.~Tang, and F.~Blaabjerg, ``On the inertia of future
  more-electronics power systems,'' \emph{IEEE Journal of Emerging and Selected
  Topics in Power Electronics}, vol.~7, no.~4, pp. 2130--2146, 2018.

\end{thebibliography}

\end{document}